\documentclass[aps,pra,twocolumn,amsmath,amssymb,showpacs,floatfix,superscriptaddress]{revtex4-1}

\usepackage{times}
\usepackage[utf8x]{inputenc}
\usepackage[english]{babel}
\usepackage[T1]{fontenc}
\usepackage{lmodern}
\usepackage{amssymb}
\usepackage{amsmath}
\usepackage{amsthm}
\usepackage[pdftex]{graphicx}
\usepackage{epstopdf}
\usepackage{braket}
\usepackage[bottom]{footmisc}
\usepackage{dcolumn}
\usepackage{bm}
\usepackage{color}
\usepackage{relsize,dsfont,mathrsfs,empheq,verbatim,upgreek}
\usepackage{etoolbox}
\usepackage[caption = false]{subfig}

\usepackage[hidelinks]{hyperref}
\usepackage{bbm}

\usepackage{enumerate}

\newcommand{\Tr}{\mathrm{Tr}}
\newcommand{\Lt}{\mathcal{L}}

\begin{document}

%%%%%%%%%%%%%%%%%%%%%%%%%%%%%%%%%%%%%%%%%%%%%

\title{Local and global approaches to the thermodynamics of pure decoherence processes in open quantum systems}

%%%%%%%%%%%%%%%%%%%%%%%%%%%%%%%%%%%%%%%%%%%%%

\author{Irene Ada Picatoste}

\affiliation{Institute of Physics, University of Freiburg, 
Hermann-Herder-Stra{\ss}e 3, D-79104 Freiburg, Germany}

\author{Alessandra Colla}

\affiliation{Dipartimento di Fisica Aldo Pontremoli, Università degli Studi di Milano, via Celoria 16, 20133 Milan, Italy}

\affiliation{ INFN, Sezione di Milano, Via Celoria 16, I-20133 Milan, Italy}

\author{Heinz-Peter Breuer}

\affiliation{Institute of Physics, University of Freiburg, 
Hermann-Herder-Stra{\ss}e 3, D-79104 Freiburg, Germany}

\affiliation{EUCOR Centre for Quantum Science and Quantum Computing,
University of Freiburg, Hermann-Herder-Stra{\ss}e 3, D-79104 Freiburg, Germany}

\begin{abstract}
We study the nonequilibrium thermodynamics of pure decoherence processes in open quantum systems coupled to a thermal reservoir.
We review various definitions of central quantities, such as internal energy, work, heat and entropy production,
developed within local and global approaches to quantum thermodynamics. Within local approaches thermodynamic quantities
only refer to the open system's degrees of freedom, while in the global approaches certain quantities are defined by referring explicitly 
to the reservoir degrees of freedom. Employing a microscopic, analytically solvable model, we perform a comparison of these two perspectives,
revealing substantial differences in the thermodynamic quantities and in the formulations of the first and second law. The main reason 
for these discrepancies is the fact that the global approaches involve the system-reservoir interaction which exchanges a large amount 
of energy with the environment, while the average open system energy is constant in time because the dynamics represents pure decoherence
and does not affect the open system populations.
\end{abstract}

\date{\today}

\maketitle

\section{Introduction}\label{sec:Intro}

Quantum thermodynamics is an emergent field which aims to study the energetics of quantum systems \cite{Gemmer2004,Binder2018}.  
Closed systems in quantum mechanics follow unitary dynamics, determined by the Hamiltonian of the system, and their average energy is defined as the expectation value of the Hamiltonian. When the Hamiltonian is time-independent, energy is conserved; when it is not, any change of the energy is considered a work contribution, since the system is closed, and therefore there cannot be any heat exchange with an external reservoir. 
When the system is coupled to an environment it is said to be an open system \cite{Breuer2002}. Open systems exhibit more complex dynamics, where unitarity is lost, which leads to many questions: What is the energy of an open system? How to identify the heat flow, the work exchange? Can system and environment exchange energy only in the form of heat, or can they also perform work on each other?

An important quantity in thermodynamics is entropy production, which aims at giving a measure for the irreversibility of a process, and can also be understood as an informational measure describing our knowledge about the system. 
In classical thermodynamics 
entropy production can be defined by means of the Clausius relation
\begin{eqnarray}
\label{eq:Clausius}
    \Sigma_S^\text{Cl}(t) = \Delta S_S(t) - \beta Q_S(t),
\end{eqnarray}
where $\Delta S_S(t)$ denotes the change of entropy of the system, $Q_S(t)$ the heat flow and $\beta$ the inverse temperature.
It is not clear yet how to extrapolate this concept to quantum systems, and a variety of approaches have been discussed in the literature \cite{Esposito2010, CollaPhD, Strasberg2021, Deffner2010, Nicacio2023, Landi2021, Rivas2020, Elouard2023}. Basically, one can classify them under two viewpoints: Those that are based solely on the reduced description of the open system, such as the weak coupling formulation \cite{Alicki1987} or the minimal dissipation proposal \cite{Colla2022a}, and can thus be characterized as \textit{local} approaches, and those which take a \textit{global} point of view and include information about the bath degrees of freedom \cite{Esposito2010}.

Here, we will study two particular approaches: The first is a local one which extends the weak coupling Spohn formulation for entropy production in terms of relative entropy \cite{Spohn1978}, employing the minimal dissipation formulation of the first law \cite{Colla2022a}. The second approach is a global one which follows 
the definition of entropy production proposed by Esposito, Lindenberg and van den Broeck \cite{Esposito2010}, including the variant of the first law described in \cite{Landi2021}. These two approaches to entropy production are fundamentally different in nature -- one takes an informational point of view and is focused on the open system distance from the steady state, while the other follows the Clausius formulation of Eq.~\eqref{eq:Clausius}, based on the discrepancy between entropy flows, and assumes the expected entropy flow due to heat to be given by energy change in the environment. First law quantities also take very different expressions, since the minimal dissipation approach defines an effective energy operator for the system based on its dynamics, while the global approach assumes all energy change in the environment is a heat flow into the system.

In this work we will carry out a detailed comparison of the above approaches to entropy production and to the first law by focusing on a particular kind
of system-environment models leading to pure decoherence dynamics (a true quantum effect), in which the coherences of the open system density matrix decay due to the build up of
system-environment correlations, while its populations stay constant in time \cite{Breuer2002}.
Because of the different dynamics on the system and on the environment sides, the approaches mentioned yield greatly differing results and, thus, pure decoherence models give valuable insights into the physical pictures provided by the different approaches to quantum thermodynamics \cite{Popovic2023, Marcantoni2017-tl}.

The article is structured as follows: In Sec.~\ref{sec:QuantumThermo} we introduce the general formalism for open systems and present the local (Sec.~\ref{sec:min_diss}) and the global (Sec.~\ref{sec:esposito_theory}) approaches. Section \ref{sec:general_dephasing_models} gives some insights into general dephasing models, and in Sec.~\ref{sec:example} a specific example is analyzed, studying both the entropy production and the second law (Sec.~\ref{sec:example_secondlaw}), and first law quantities (Sec.~\ref{sec:example_firstlaw}). Finally, a discussion of the results and conclusions are given in Sec.~\ref{sec:conclusions}.

\section{Quantum Thermodynamics of Open Systems}\label{sec:QuantumThermo}

We consider an open quantum system $S$ and an environment $E$, which are coupled such that the total microscopic Hamiltonian is given by 
\begin{eqnarray}\label{eq:gen_Ham}
    H(t) = H_S(t) + H_E + H_I(t).
\end{eqnarray}
Here, $H_S(t)$ is the (possibly time dependent) system Hamiltonian, $H_E$ is the Hamiltonian of the environment and $H_I(t)$ describes the interaction between them \cite{Breuer2002}. The global system is assumed to be closed, and thus its state, described by the density matrix $\rho_{SE}(t)$, follows unitary evolution according to von Neumann's equation $\dot{\rho}_{SE} (t) = -i \left[ H(t), \rho_{SE}(t) \right]$ (setting $\hbar = 1$). When interested in a reduced description of the system's degrees of freedom alone we take the partial trace over the environmental degrees of freedom to obtain the open system density matrix
\begin{eqnarray}
    \rho_S(t) = \Tr_E \{ \rho_{SE}(t)\}.
\end{eqnarray}
We further assume that system and environment start in an uncorrelated state 
\begin{eqnarray}
    \rho_{SE} (0) = \rho_S(0) \otimes \rho_E(0),
\end{eqnarray}
although extensions to take into account initial correlations have been studied \cite{Colla2022b, Vacchini2016-se, Alipour2020}. 
With this setup in mind, we will now go on to describe two different approaches to the study of quantum thermodynamics.

\subsection{Local approach}\label{sec:min_diss}
We first consider a local approach to the thermodynamics of an open system, which is based only on knowledge of the system and its evolution in time. It stems from the viewpoint that the bath is experimentally very hard to keep track of, as it generally consists of a very large or infinite number of degrees of freedom, and what one typically has access to are only the degrees of freedom of the system.

The dynamics of the reduced system can be written in the most general form as a time-convolutionless (TCL) master equation \cite{Shibata1977, Shibata1979} in generalized Lindblad form \cite{Hall2014Apr}:
\begin{equation} \label{tcl-meq}
 \frac{d}{dt}\rho_S(t) = \Lt_t[\rho_S(t)] = -i \left[K_S(t),\rho_S(t)\right] + {\mathcal{D}}_t  [\rho_S(t)].
\end{equation}
It comprises a commutator part with a Hamiltonian $K_S(t)$, and a dissipator part which has the following structure: 
\begin{equation} \label{dissipator}
 {\mathcal{D}}_t [\rho_S] = \sum_{k}\gamma_{k}(t)\Big[L_{k}(t)
  \rho_S L_{k}^{\dag}(t) - \frac{1}{2}\big\{L_{k}^{\dag}(t)L_{k}(t),\rho_S\big\}\Big]
\end{equation}
with time dependent rates $\gamma_k(t)$ and Lindblad operators $L_k(t)$. The rates can become temporarily negative \cite{Breuer1999b,Breuer2002}, signifying the violation of CP divisibility and representing a necessary condition for information backflow and quantum non-Markovianity \cite{Breuer2016a}.

We will start by looking at the possible definitions for entropy production from this open system point of view, using the idea that the information available is the generator of the dynamics in Eq.~\eqref{tcl-meq}. This is broader than just having access to the state of the reduced system at a certain time, since some information about the bath is encoded in the generator, but still all information needed is available by just keeping track of the system's degrees of freedom. 

Our proposed definition of entropy production takes thus an informational point of view, extending the weak coupling formulation of Spohn \cite{Spohn1978}. This formulation has already been studied in certain scenarios \cite{Strasberg2019}, and our aim here is to use it for arbitrary quantum systems. Assuming there always exists a unique state  $\rho_S^\star (t)$ constituting an instantaneous fixed point (IFP) of the dynamics, such that 
\begin{equation}\label{eq:IFP_def}
    \mathcal{L}_t \left[\rho_S^\star (t) \right] = 0,
\end{equation} 
we define entropy production rate $\sigma_S(t)$ as: 
\begin{equation}\label{eq:entropy_prod_rate_spohn}
    \sigma_S^\text{loc} (t) = - \left. \frac{d}{d\tau} \right\vert_{\tau = 0} D\left(\rho_S(t + \tau) \vert \vert \rho_S^\star (t) \right) ,
\end{equation}
where $D$ is the relative entropy  defined as 
\begin{equation}
D(\rho \vert \vert \sigma) = \Tr \{\rho \log \rho\} - \Tr \{\rho \log \sigma \}.
\end{equation}
Special dynamics where the IFP is not unique, but all of them give the same entropy production rate will be discussed in Sec.~\ref{sec:general_dephasing_models}.

The quantity defined by Eq.~\eqref{eq:entropy_prod_rate_spohn} evaluates the rate of change of the distance between the actual state and the IFP of the dynamics at each instant in time. When the dynamics is driving the system towards the local fixed point, entropy production rate is positive, making the produced entropy increase. However, the interaction with the environment can also drive the system away from the IFP, which yields a negative entropy production rate. Thus, the quantity \eqref{eq:entropy_prod_rate_spohn} provides a measure for the irreversibility of the process, which can also be linked under certain conditions to the degree of non-Markovianity (memory effects) of the dynamics \cite{Colla2022a,CollaPhD}.

More explicitly, Eq.~\eqref{eq:entropy_prod_rate_spohn} can be rewritten as 
\begin{eqnarray}\label{eq:ent_prod_rate_md_explicit}
    \sigma_S^\text{loc} (t) &=& - \Tr \left\{ \dot{\rho}_S(t) \log \rho_S(t) \right\} + \Tr \left\{ \dot{\rho}_S(t) \log \rho_S^\star (t) \right\} \nonumber  \\
    &=& \dot{S}_S(t) + \Tr \left\{ \dot{\rho}_S(t) \log \rho_S^\star (t) \right\} ,
\end{eqnarray}
with $S_S(t) = - \Tr \{ \rho_S(t) \log \rho_S(t) \}$ the von Neumann entropy of the system. This formulation is more reminiscent of the Clausius formulation in Eq. \eqref{eq:Clausius}, but still the second term in Eq.~\eqref{eq:ent_prod_rate_md_explicit} is not necessarily the heat flow. We will use this expression later to see that the local approach reconciles with the Clausius formulation in some special cases, and modifies it in others.
Integrating over time, we get the entropy production given by 
\cite{Deffner2010}
\begin{eqnarray}
    \Sigma_S^\text{loc}(t) &=&  \int_0^t  d\tau \; \sigma_S^\text{loc}(\tau)  \nonumber \\
    &=& \Delta S_S(t) + \int_0^t d\tau \Tr \{\dot{\rho}_S (\tau) \log \rho_S^\star (\tau) \} \nonumber\\
    &=& D(\rho_S(0) \vert \vert \rho_S^\star (0)) -   D(\rho_S(t) \vert \vert \rho_S^\star (t))  \nonumber\\
    && - \int_0^t d\tau \Tr \{ \rho_S(\tau) \partial_\tau \log \rho_S^\star(\tau) \}.
\end{eqnarray} 

Secondly, we are also interested in the first law of thermodynamics, and the definitions of internal energy, work, and heat in an open system. In order to find these quantities we turn again to the master equation governing the dynamics of the system in Eq.~\eqref{tcl-meq}. What we notice is that the commutator in Eq.~\eqref{tcl-meq} resembles the von Neumann equation, and can describe the coherent part of the dynamics, while the dissipator structure is what allows for all open systems effects: dissipation, decoherence, irreversibility or backflow of information. However, the splitting of the master equation into Hamiltonian and dissipator parts is highly non-unique, as there exist several transformations which leave the equation invariant while shifting Hamiltonian and dissipator contributions \cite{Breuer2002, Parthasarathy1992}. The dissipator structure is  more general, to the point that the whole Hamiltonian can be completely rewritten in dissipator form \cite{Hayden2022May}, while the reverse is not true: the commutator is a special structure which can only describe the coherent part of the dynamics. 

The principle of minimal dissipation \cite{Hayden2022May, Colla2022a} prescribes a unique splitting by minimizing the dissipator part with respect to a certain superoperator norm. The physical intuition behind this amounts to writing everything that can be seen as giving rise to coherent dynamics into the Hamiltonian, and leaving only what remains as a dissipator. Thus, an effective time-dependent Hamiltonian $K_S(t)$ emerges from the dynamics.  
The minimization procedure has been proven to be equivalent to choosing traceless Lindblad operators \cite{Hayden2022May}, which provides a straightforward way of writing a master equation in minimal dissipation form. Additionally, for any orthonormal basis of states $\{ |n \rangle \}_{n=1,...,N}$, with $N$ the dimension of the Hilbert space, the effective Hamiltonian can be found from the generator by computing \cite{CollaPhD, Colla2025} 
\begin{eqnarray}\label{eq:KS_general}
    K_S(t) = \frac{1}{2 i N} \sum_{m, n} \Bigl[|n \rangle \langle m |, \mathcal{L}_t \bigl[|m \rangle \langle n | \bigr]\Bigr].
\end{eqnarray}

This emergent Hamiltonian is then taken as an effective energy operator for the open system. Internal energy is defined as 
\begin{eqnarray}
    U_S^{\text{loc}} (t) = \langle K_S(t) \rangle_t = \text{Tr} \{ K_S(t) \rho_S(t) \},
\end{eqnarray}
which describes the energy of the system at each point in time when in interaction with the environment. 
Analogously to what is done for weak coupling thermodynamics \cite{Alicki1979May}, work and heat are defined as the integrated energy change due to a change in the Hamiltonian (work) or due to a change in the state of the system (heat): 
\begin{eqnarray}
 W_S^{\text{loc}}(t) &=& \int_0^t d\tau \, \Tr \big\{ \dot{K}_S(\tau) \rho_S(\tau) \big\},  \label{work_md} \\
 Q_S^{\text{loc}}(t) &=& \int_0^t d\tau \, \Tr \big\{ K_S(\tau) \dot{\rho}_S(\tau)  \big\}.  \label{heat_md}
\end{eqnarray}
A particular feature of this approach is that the emergent Hamiltonian $K_S(t)$ can be time dependent even if the microscopic Hamiltonian is not \cite{Colla2025-ne}, which means that system and environment can exchange energy not only in the form of heat, but also in the form of work. 
This prescription results in the following formulation for the first law of thermodynamics:
\begin{eqnarray}\label{first-law_md}
 \Delta U_S^{\text{loc}} (t) \equiv U_S^{\text{loc}}(t) - U_S^{\text{loc}}(0) = W_S^\text{loc}(t) + Q_S^\text{loc}(t).
\end{eqnarray}

Having established a definition for heat allows us to relate the entropy production rate in Eq. \eqref{eq:ent_prod_rate_md_explicit} with the typical Clausius expression.
Indeed, in some scenarios (e.g. at weak coupling with the environment initially in a thermal state at temperature $T$) the effective Hamiltonian reduces to (or can be approximated by) the bare Hamiltonian of the system, $K_S(t) = H_S$ and, when a Gibbs state of this Hamiltonian with the inverse temperature of the environment $\beta = 1 / T$ is a fixed point of the dynamics \cite{Davies1974, Spohn1978}, then: 
\begin{eqnarray}
    \rho_S^\star = \frac{e^{- \beta H_S}}{Z_S},
\end{eqnarray}
where $Z_S = \Tr \{ e^{- \beta H_S}\}$ is the partition function. 
Then we recover the original Clausius formulation 
\begin{eqnarray} \label{eq:Cl_local_weak}
    \sigma_S^\text{loc} (t) = \dot{S}_S(t) - \beta \dot{Q}_S^\text{loc} (t).
\end{eqnarray}
Outside of weak coupling we can still find generalized detailed balance scenarios \cite{CollaPhD, Colla2024Dec} where the IFP can be written as a Gibbs state of the effective Hamiltonian with some time-dependent renormalized temperature $\beta_r(t)$: 
\begin{eqnarray}
    \rho_S^\star (t) = \frac{e^{- \beta_r(t) K_S(t)}}{Z_S^\star},
\end{eqnarray}
where $Z_S^\star = \Tr \{e^{- \beta_r (t) K_S(t) } \}$.
Then, a modified Clausius expression for entropy production results from Eq. \eqref{eq:ent_prod_rate_md_explicit}:
\begin{eqnarray}
    \sigma_S^\text{loc} (t) = \dot{S}_S(t) - \beta_r(t) \dot{Q}_S^\text{loc}(t),
\end{eqnarray}
where now the temperature factor multiplying the heat current is substituted by the renormalized temperature.

On the other hand, one could also take an a priori Clausius formulation for entropy production and impose the entropy production in Eq. \eqref{eq:Cl_local_weak}, 
even when it differs from Eq.~\eqref{eq:entropy_prod_rate_spohn}:
\begin{eqnarray}\label{eq:Clausius_local}
    \sigma_S^\text{loc, Cl} (t) = \dot{S}_S(t) - \beta \dot{Q}_S^\text{loc}(t),
\end{eqnarray}
with $\beta$ the initial temperature of the environment \cite{Colla2022a}. This formulation is different from the informational one in Eq. \eqref{eq:entropy_prod_rate_spohn}, but we will see in Sec. \ref{sec:general_dephasing_models} that they both give the same result for pure decoherence dynamics.

\subsection{Global approach}\label{sec:esposito_theory}

The other approach to quantum thermodynamics is one that takes a global perspective, and considers the evolution of the whole system+environment. First introduced in \cite{Esposito2010} for systems coupled to a thermal bath, it is based on a formulation for the second law and heat, while expressions for internal energy, work, and the first law are still under debate \cite{Esposito2010, Landi2021}. In general, this global approach takes a different point of view from the local one, since it requires some knowledge about the environment, in addition to the knowledge about the system, in order to calculate thermodynamic quantities, rather than restricting our knowledge to the evolution of the degrees of freedom of the system only. 

The initial state of system+environment is taken to be 
\begin{eqnarray}
    \rho_{SE} (0) = \rho_S (0) \otimes \rho_E^\text{eq} = \rho_S(0) \otimes \frac{1}{Z_E} e^{- \beta H_E},
\end{eqnarray}
with the environment in a Gibbs state described by $\beta = (k_B T)^{-1}$ the inverse temperature, and with $Z_E = \Tr \{e^{-\beta H_E}\}$ the partition function. 

In order to obtain a second law of thermodynamics, the global approach follows Clausius's formulation, and expresses entropy production during the process as the difference between the change of von Neumann entropy in the system and the change of entropy expected from the heat flow:
\begin{eqnarray}\label{eq:ent_prod_ELB}
    \Sigma_S^\text{gl} (t) = \Delta S_S(t) - \beta Q_S^\text{gl} (t) ,
\end{eqnarray}
using $\beta$ the initial temperature of the environment.

Thus, a definition for heat is needed. The idea behind this approach is that any change in the internal energy of the bath will be a heat contribution to the system. Therefore, the heat exchanged with the system during the process is minus the change of energy of the environment: 
\begin{eqnarray}\label{eq:heat_ELB}
    Q_S^\text{gl} (t) =  \langle H_E \rangle_0 - \langle H_E \rangle_t = - \int_0^t d\tau \Tr \{H_E \dot{\rho}_{SE} (\tau) \} .
\end{eqnarray}
Due to the unitarity of the global evolution (implying $\Tr \{ (H(t) \dot{\rho}_{SE}(t) \} = 0)$, we can also write heat as
\begin{eqnarray}\label{eq:heat_ELB_2}
    Q_S^\text{gl}(t) = \int_0^t d\tau \Tr \{(H_S(\tau)+H_I(\tau)) \dot{\rho}_{SE} (\tau) \}.
\end{eqnarray}
This expression for heat allows us to rewrite Eq. \eqref{eq:ent_prod_ELB} as
\begin{eqnarray}
    \Sigma_S^\text{gl} (t) = D\left(\rho_{SE} (t) \vert \vert \rho_S(t) \otimes \rho_E^\text{eq}\right).
\end{eqnarray}
The latter formulation can illustrate the informational aspect of Eq. \eqref{eq:ent_prod_ELB}, since it compares the actual state of system+environment at time $t$ and the uncorrelated state constructed by taking the tensor product between the reduced state of the system $\rho_S(t)$ and the equilibrium state of the reservoir $\rho_E^\text{eq}$ (also the state at which the environment is initialized). In reality, the state $\rho_{SE} (t)$ will build up correlations between system and environment throughout the process, but the tensor product between the state of the system (which we have access to) and a thermal state of the environment is our best approximation when unable to follow the dynamics of the environment. This is why the entropy production $\Sigma_S^\text{ELB} (t)$ can be seen as the information lost throughout the evolution. 

The entropy production defined in Eq. 
\eqref{eq:ent_prod_ELB} and the heat in Eq. \eqref{eq:heat_ELB} still allow for different ways in which to define internal energy and work from the global perspective. 
One way (introduced in \cite{Esposito2010} by Esposito, Lindenberg and van den Broeck, and denoted hereon as ELB) considers that, since the global system is closed, any change of its energy will be a work contribution, and the fact that the environment Hamiltonian $H_E$ is time independent is reason enough to say that the total work exchange is work done on the open system. Hence, we obtain the following expression: 
\begin{eqnarray}\label{eq:work_ELB}
    W_S^\text{gl,ELB} &=& \langle H(t) \rangle_t - \langle H(0)\rangle_0 \nonumber \\
    &=& \int_0^t d\tau \Tr \; \{ (\dot{H}_S(\tau) + \dot{H}_I (\tau) ) \rho_{SE} (\tau) \}.
\end{eqnarray}

Adding the expressions for heat \eqref{eq:heat_ELB_2} and work \eqref{eq:work_ELB}, we get the following natural definition of internal energy:
\begin{equation}
    \label{eq:U_ELB}
    U_S^\text{gl,ELB} (t) = \Tr \{ (H_S(t) + H_I(t)) \rho_{SE} (t) \},
\end{equation}
from which we can recover the first law of thermodynamics
\begin{eqnarray}
    \label{eq:first_law_ELB}
    \Delta U_S^\text{gl,ELB} = W_S^\text{gl,ELB} + Q_S^\text{gl}.
\end{eqnarray}
This approach can be regarded as equivalent to taking an effective Hamiltonian $H_\text{eff}(t) = H_S(t) + H_I(t)$ to define first law quantities, where the whole interaction Hamiltonian is considered as part of the energy of the open system.  

Another way (discussed by Landi and Paternostro in \cite{Landi2021}, and denoted here by LP) is to still consider the weak coupling formulation for internal energy, using the bare Hamiltonian of the system: 
\begin{eqnarray}
    U_S^\text{gl,LP} (t)  = \Tr \{ H_S \rho_S(t) \}.
\end{eqnarray}
To recover the first law $\Delta U_S^\text{gl,LP}(t) = W_S^\text{gl,LP}(t) + Q_S^\text{gl}(t)$ the work contribution is 
\begin{eqnarray}
    \dot{W}_S^\text{gl,LP} = \Tr \{ \dot{H}_S(t) \rho_S(t) \} - \Tr \{H_I(t) \dot{\rho}_S(t) \}.
\end{eqnarray}

Altogether, the global approach keeps track of the energy of the environment and ascribes any energy change of the bath to a heat contribution to the energy of the system. It then follows the Clausius formulation for entropy production, defining it as the mismatch between the change of von Neumann entropy of the system and the entropy change attributed to the heat flow. Finally, internal energy can be defined using the bare system Hamiltonian (LP), or joining the system and environment Hamiltonians (ELB), and a corresponding definition for work follows from the first law of thermodynamics. 

\section{General decoherence models}
\label{sec:general_dephasing_models}
The generator for the master equation of an $N$ level system can be written in the most general form as 
\begin{eqnarray}\label{eq:general_generator}
    \mathcal{L}_t \left[ \rho_S(t) \right] = \sum_{jkln} \gamma_{jkln} (t) |j\rangle \langle k | \rho_S(t) |l \rangle \langle n|,
\end{eqnarray}
where the coefficients $\gamma_{jkln}(t)$ depend on the choice of the orthonormal basis $\{|j\rangle\}$ of the Hilbert space of the system. 
In this work we want to study pure decoherence processes. These are described by a microscopic Hamiltonian of the form of Eq. \eqref{eq:gen_Ham} with an interaction Hamiltonian which commutes with the system Hamiltonian: 
\begin{eqnarray}
    \left[ H_S(t), H_I(t) \right] = 0.
\end{eqnarray}
This property leads to a symmetry represented by a unitary $U_\lambda = e^{i \lambda S}$, for any operator $S$ such that $\left[H_S(t), S \right] = 0$. Since the microscopic Hamiltonian is invariant under the symmetry, the generator of the dynamics will also be: 
\begin{eqnarray}
\label{eq:symmetry}
   && U_\lambda H U_\lambda^\dagger = H \quad \forall \; \lambda \nonumber \\
   && \Rightarrow U_\lambda \mathcal{L}_t \left[ \rho_S(t) \right] U_\lambda^\dagger = \mathcal{L}_t \left[ U_\lambda \rho_S(t) U_\lambda^\dagger \right] 
\end{eqnarray}
Applying this symmetry to the generator in Eq. \eqref{eq:general_generator} results in the following form for the generator of any pure  decoherence process \cite{Chruscinski2022}:
\begin{eqnarray}\label{eq:generator_decoherence}
    \mathcal{L}_t \left[ \rho_S(t) \right] = \sum_{jk} \gamma_{jk} (t) |j\rangle \langle j | \rho_S(t) |k \rangle \langle k |,
\end{eqnarray}
where the coefficients $\gamma_{jk}(t)$ fulfill the conditions
\begin{eqnarray}\label{eq:conditions_e_coeffs}
    \gamma_{jk} (t)= \gamma_{kj}^*(t) \equiv \gamma_{jjkk}(t), \quad \gamma_{jj} (t)= 0,
\end{eqnarray}
and $\{ |j\rangle \}$ is the eigenbasis of the bare system Hamiltonian $H_S(t)$. The details of this derivation can be found in Appendix \ref{app:general_dephasing_ME}.

The master equation allows us to calculate entropy production in the local approach. We can verify that any diagonal operator in the energy eigenbasis of the bare system Hamiltonian is a fixed point of the dynamics: 
\begin{eqnarray}
    \rho_S^\star (t) = \sum_\alpha p_\alpha^\star |\alpha \rangle \langle \alpha | \hspace{0.5cm}\Rightarrow \mathcal{L}_t \left[ \rho_S^\star (t) \right] = 0
\end{eqnarray}
Using this fixed point we find entropy production rate to be (thanks to Eq. \eqref{eq:conditions_e_coeffs}) 
\begin{eqnarray}
    \sigma_S^\text{loc} (t) &=&  \dot{S}_S(t) + \Tr \{\mathcal{L}_t \left[\rho_S(t) \right] \ \log \rho_S^\star (t) \} \nonumber\\
    &=&  \dot{S}_S(t) + \sum_{jk\alpha} \gamma_{jk} (t)\log p_\alpha^\star (t) \delta_{jk} \delta_{j \alpha} \rho_S^{\alpha \alpha}(t) \nonumber \\
    &=& \dot{S}_S(t).
\end{eqnarray}

We can additionally calculate the alternative local entropy production following the Clausius formulation (see Eq. \eqref{eq:Clausius_local}). For this we need the heat contribution, and therefore we have to start by finding the effective Hamiltonian. Using Eq. \eqref{eq:KS_general} we find 
\begin{eqnarray}
    K_S(t) = \frac{1}{N} \sum_n \left( \sum_m \Im (\gamma_{mn}(t)) \right) |n\rangle \langle n|,
\end{eqnarray}
and thus the effective Hamiltonian of pure decoherence processes is diagonal in the energy eigenbasis of the bare system. Then, following Eq. \eqref{heat_md} we find:
\begin{eqnarray}
    \dot{Q}_S^\text{loc} (t) &=& \Tr \{K_S(t) \mathcal{L}_t \left[\rho_S(t) \right] \} \nonumber \\
    &=& \frac{1}{N} \sum_{nmjk} \Im (\gamma_{nm}(t)) \gamma_{jk}(t) \delta_{jk} \delta_{jn} \rho_S^{nn}(t) \nonumber \\
    &=& 0.
\end{eqnarray}

Hence, we see that any local formulation for entropy production predicts an entropy production equal to the change of von Neumann entropy of the system, without any entropy flow expected from the environment.

On the other hand, the global approach requires information about the bath, and the entropy production of the decoherence process will depend heavily on the coupling chosen and the properties of the bath. This will be seen better in the following example. 

\section{Example: two level system coupled to a bosonic reservoir}\label{sec:example}

We look now at a simple integrable dephasing model to provide an explicit example and illustrate the findings before. 
In order to do so, we study a spin (which will be our open system of interest) linearly coupled to a bosonic reservoir (the environment). The total system is described by the following microscopic Hamiltonian \cite{Breuer2002, Morozov_2012}: 
\begin{eqnarray} \label{total-ham}
 H &=& H_S + H_E + H_I \\
 &=& \frac{\omega_0}{2} \sigma_z + \sum_k \omega_k  b_k^{\dagger}b_k
 + \sigma_z \otimes \sum_k \left( g_k b_k^{\dagger} + g^*_k b_k \right). \nonumber
\end{eqnarray}
Here $\omega_0$ is the bare frequency of the system and $\omega_k$ the frequencies of the modes in the environment with creation and annihilation operators $b_k^\dagger, b_k$, which satisfy the commutation relations $[b_k, b_{k^\prime}^\dagger ] = \delta_{k k^\prime}$. The interaction couples system and environment with $g_k$ the coupling constant between the system and mode $k$.

At the initial time of the evolution we take system and environment to be in an uncorrelated state 
\begin{equation} \label{initial_state}
 \rho_{SE}(0) = \rho_S(0) \otimes \rho_E(0),
\end{equation}
with the environment initially in a Gibbs state
\begin{equation} \label{env_gibbs}
 \rho_E(0) = \rho_E^{eq} = \frac{1}{Z_E} e^{-\beta H_E}.
\end{equation}

In this model the system Hamiltonian commutes with the interaction Hamiltonian $\left[H_S, H_I \right] = 0$, hence giving rise to pure dephasing or pure decoherence dynamics on the system side. This results in the fact that the Heisenberg-picture operator $\sigma_z(t)$ will not change in time, leaving the populations of the two-level system constant, while we can expect the coherences of the system to decay in time. 

On the other hand, there is not only dephasing on the environment side, as the interaction operator on the Hilbert space of the environment $B = \sum_k \left( g_k b_k^\dagger + g_k^* b_k \right)$ does not commute with the Hamiltonian of the environment $H_E$. This constitutes an interesting case study that we look at here from a thermodynamic perspective. 

Since the model is integrable, we can get analytical expressions for the evolution of the degrees of freedom of system and environment at each point in time. Additionally, we can construct an exact TCL master equation for the dynamics of the reduced system. We will now present the solution to the model and the derivation of the master equation, and will then use them to calculate thermodynamic quantities according to the two different approaches
discussed in Sec.~\ref{sec:QuantumThermo}.

\subsection{Solution of the model and master equation}

The model introduced in Sec. \ref{sec:example} is simple enough to present an analytical solution. Following the derivation given in \cite{Breuer2002} (for details see Appendix \ref{app:calculations_rho_s}), we find the dynamics for the reduced density matrix of the system. 

Each element of the system density matrix in the energy eigenbasis of the bare system is obtained by 
\begin{eqnarray}
    \rho_{S}^{ij} (t)= \langle i | \rho_S(t) | j \rangle,
\end{eqnarray}
where $\sigma_z |0 \rangle = - |0\rangle$, $\sigma_z |1\rangle = |1\rangle$. We obtain the following: 
\begin{eqnarray}
    \rho_S^{00}(t) &=& \rho_S^{00}(0), \label{eq:rho_S_00} \\
    \rho_S^{11}(t) &=& \rho_S^{11}(0) , \label{eq:rho_S_11} \\
    \rho_S^{01}(t) &=& \rho_S^{01}(0) e^{i \omega_0 t} e^{-\eta(t)}, \label{eq:rho_S_01} \\
    \rho_S^{10}(t) &=& \rho_S^{10}(0) e^{-i \omega_0 t}  e^{- \eta(t)}. \label{eq:rho_S_10}
\end{eqnarray}
The populations indeed do not change in time, while the coherences oscillate with the bare frequency of the system and decay with a decoherence function $\eta(t)$. This function is defined as 
\begin{eqnarray}
    \eta(t) = 2 \int_0^\infty d\omega J(\omega) \frac{1-\cos(\omega t)}{\omega^2} \coth\left(\frac{\beta \omega}{2}\right),
\end{eqnarray}
where we have introduced the spectral density $J(\omega)$ containing the information about the modes in the bath and how strongly they each couple to the system:
\begin{eqnarray}\label{eq:J_def}
    J(\omega) = \sum_k |g_k|^2 \delta (\omega - \omega_k).
\end{eqnarray}
The decoherence function $\eta (t)$ is real and always positive, and depends on both the spectral density and the initial temperature of the environment $\beta$. 

In order to apply the principle of minimal dissipation, we need to write down the exact, TCL master equation for the system. This can be done by using the exact solution for the dynamics of the system found in Eqs. \eqref{eq:rho_S_00}-\eqref{eq:rho_S_10}. Again, the details of the procedure are given in Appendix \ref{app:master_equation}, where we find the exact master equation
\begin{eqnarray}\label{eq:master_equation_example}
    \frac{d}{dt} \rho_S(t) = -i \left[H_S, \rho_S(t) \right] + \Gamma(t) \left[\sigma_z \rho_S(t) \sigma_z - \rho_S(t)\right]
\end{eqnarray}
with the bare system Hamiltonian in the commutator and a dissipator with Lindblad operator $\sigma_z$ and a time dependent rate
\begin{eqnarray}
    \Gamma(t) = \frac{\dot{\eta} (t)}{2} = \int_0^\infty d\omega J(\omega) \frac{\sin(\omega t)}{\omega} \coth\left(\frac{\beta \omega}{2} \right).
\end{eqnarray}
Eq. \eqref{eq:master_equation_example} is already in minimal dissipation form, since $\sigma_z$ is a traceless operator. The master equation and the dynamics of the system are all that is needed to calculate the thermodynamic quantities within the local approach. 

Finally, we need the expectation value of the system and interaction Hamiltonians at each point in time to calculate the thermodynamic quantities in the global approach. The evolution of the expectation value of the interaction Hamiltonian can also be found in this model, and turns out to be (see details in Appendix \ref{app:calculations_HI}):
\begin{eqnarray}
    \langle H_I \rangle_t=  - 2 \int_0^\infty d\omega J(\omega) \frac{1- \cos(\omega t)}{\omega}.
\end{eqnarray}

\subsection{Comparison of thermodynamic quantities}

\subsubsection{Entropy production}
\label{sec:example_secondlaw}

We start by looking at the entropy production rate and the second law of thermodynamics, from both the local and the global approaches. 

For the local approach we need to find the instantaneous fixed point of the master equation \eqref{eq:master_equation_example}. This master equation does not have a unique IFP, but infinitely many: Any state which is diagonal in the energy eigenbasis will be an IFP, independently of the value of the populations. We can write any state that fulfills Eq. \eqref{eq:IFP_def} in the following way: 
\begin{eqnarray}
    \rho_S^\star (t) = \frac{1}{Z_S^\star} e^{- c (t) \sigma_z}, \hspace{1cm} \forall c(t) \in \mathbb{R},
\end{eqnarray}
with $Z_S^\star = \Tr \{ e^{-c(t) \sigma_z}\}$. However, every IFP will result in the same expression for entropy production. Following Eq. \eqref{eq:ent_prod_rate_md_explicit} we realize that $\log \rho_S^\star (t) = - c (t)\sigma_z - \log Z_S^\star$, such that
\begin{eqnarray}
    \sigma_S^\text{loc} (t) &=& - \Tr \{ \dot{\rho}_S(t) \log \rho_S(t) \} \nonumber \\
    & \; & - c(t) \; \Tr \{\sigma_z \dot{\rho}_S(t) \} - \log Z_S^\star \; \Tr\{\dot{\rho}_S(t) \} \nonumber \\
    &=& - \Tr \{ \dot{\rho}_S(t) \log \rho_S(t) \} =  \dot{S}_S(t),
\end{eqnarray}
and entropy production rate is equal to the rate of change of von Neumann entropy alone (this agrees with what was found in Sec. \ref{sec:general_dephasing_models}). Entropy production in the local approach is thus given by
\begin{eqnarray}\label{eq:ent_prod_md_example}
    \Sigma_S^\text{loc} (t) = \Delta S_S(t). 
\end{eqnarray}

On the other hand, to calculate entropy production within the global approach we need to calculate the heat contribution (recall that the global approach is based on the Clausius formulation for entropy production). Following Eq. \eqref{eq:heat_ELB_2}, we can calculate the heat (i.e., the change of energy of the bath) to be 
\begin{eqnarray}
    Q_S^\text{gl} (t) = - 2 \int_0^\infty d\omega J(\omega) \frac{1-\cos(\omega t)}{\omega}.
    \label{eq:heat_ELB_example}
\end{eqnarray}
Then, using Eq. \eqref{eq:ent_prod_ELB} we find an entropy production equal to 
\begin{eqnarray}
    \Sigma_S^\text{gl} (t) = \Delta S_S(t) + 2 \beta \int_0^\infty d\omega J(\omega) \frac{1-\cos(\omega t)}{\omega}.
\end{eqnarray}

We can already see how the two approaches to entropy production differ from each other. On the one hand, the local approach gives an entropy production equal to just the change of von Neumann entropy of the system. Since any initial state will just loose its coherences, entropy production in this case is limited by the initial state of the system, and is independent of the parameters of the bath. On the other hand, the global entropy production is strongly dependent on the initial temperature of the bath $\beta$ and the spectral density $J(\omega)$. 

To get a better grasp of this result we now introduce a specific spectral density, which we choose to be Ohmic with an exponential cutoff: 
\begin{eqnarray}\label{eq:J_ohmic}
    J(\omega) = \alpha \omega e^{- \omega / \Omega}.
\end{eqnarray}
Here, $\alpha$ is a dimensionless coupling constant, which is of second order in the interaction Hamiltonian, and $\Omega$ is the frequency cutoff of the bath. Then, the heat contribution in the global approach reads
\begin{eqnarray}\label{eq:heat_ELB_example_explicit}
    Q_S^\text{gl} (t) = - 2 \alpha \Omega \frac{(\Omega t)^2}{1 + (\Omega t)^2},
\end{eqnarray}
and entropy production becomes: 
\begin{eqnarray}\label{entropy-production-global}
    \Sigma_S^\text{gl} = \Delta S_S(t) + 2 \beta \alpha \Omega \frac{(\Omega t)^2}{1 + (\Omega t)^2}.
\end{eqnarray}

\begin{figure}
    \centering
    \includegraphics[width=0.99\linewidth]{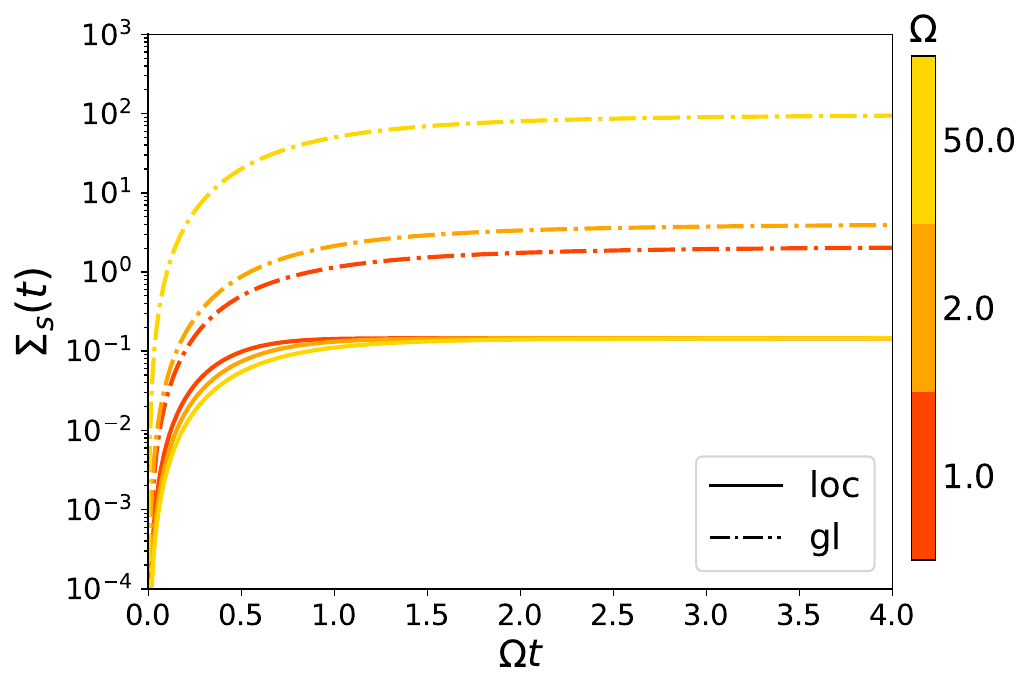}
    \caption{Entropy production in the local and global approaches as a function of time and for different values of the frequency cutoff $\Omega$. The parameters chosen are: $\omega_0 = 1.0$, $\rho_S^{11}(0) = 0.75$, $\rho_S^{01}(0) = 0.25$, $\alpha = 1.0$, $\beta = 1.0$. All parameters with dimension of energy are given in terms of an arbitrary energy unit. }
    \label{fig:second_law}
\end{figure}

The entropy production from both approaches is plotted in Fig. \ref{fig:second_law} for different values of the cutoff $\Omega$. We see that entropy production increases in both cases as the system decoheres, but in different amounts. While the final entropy $\Sigma_S^\text{loc}(t\rightarrow \infty)$ produced in the local approach coincides with the change of von Neumann entropy $\Delta S_S(t \rightarrow\infty)$ and always saturates at the same value, in the global approach one adds the heat contribution, leading to the final value $\Sigma_S^\text{gl} (t \rightarrow \infty) = \Delta S_S (t\rightarrow \infty) + 2 \beta \alpha \Omega$ for the entropy production which increases for higher values of $\Omega$.

\subsubsection{First law of thermodynamics}
\label{sec:example_firstlaw}

Next, we look at internal energy, heat and work.
For the local approach we need the effective Hamiltonian. Since the master equation \eqref{eq:master_equation_example} is already in minimal dissipation form, we directly see that the effective Hamiltonian turns out to be the bare Hamiltonian of the system without any modification:
\begin{eqnarray}
    K_S(t) = H_S = \frac{\omega_0}{2} \sigma_z.
\end{eqnarray}
The internal energy of the system when coupled to the reservoir is therefore given by the expectation value of the bare system Hamiltonian: 
\begin{eqnarray}\label{eq:U_example_md}
    U_S^\text{loc}(t) &=& \Tr \{ H_S \rho_S(t) \} = \frac{\omega_0}{2} \langle \sigma_z \rangle_t \nonumber \\
    &=& \frac{\omega_0}{2} (\rho_S^{11}(0) - \rho_S^{00}(0)),
\end{eqnarray}
and, thus, the change of internal energy is always zero:
\begin{eqnarray}\label{U_local}
    \Delta U_S^\text{loc} = 0.
\end{eqnarray}
The fact that the effective Hamiltonian is time independent implies a zero work contribution, as can be seen from Eq. \eqref{work_md} and, because of population invariance, there is also a zero heat contribution: 
\begin{eqnarray}
    W_S^\text{loc} &=& 0, \label{eq:work_md_example}\\
    Q_S^\text{loc} &=& 0. \label{eq:heat_md_example}
\end{eqnarray}
Altogether, the minimal dissipation approach gives us a constant internal energy with no heat or work exchange between system and environment. 

On the other hand, we have to consider the two different definitions for internal energy and work in the global approach (while keeping the heat studied in the previous section). 
Following the ELB approach, the change of internal energy is given by 
\begin{eqnarray}\label{eq:U_example_ELB}
    \Delta U_S^\text{gl,ELB} (t) &=&  \langle H_S + H_I \rangle_t - \langle H_S + H_I \rangle_0 \nonumber \\
    &=& - 2 \int_0^\infty d\omega J(\omega) \frac{1-\cos(\omega t)}{\omega} \nonumber \\
    &=& - 2 \alpha \Omega \frac{(\Omega t)^2}{1 + (\Omega t)^2},
\end{eqnarray}
having substituted the spectral density \eqref{eq:J_ohmic} in the last line. 
This amount coincides with the heat contribution, and we can verify that due to the time-independence of the system and the interaction Hamiltonians there is no work contribution to the change of internal energy (see definition \eqref{eq:work_ELB}):
\begin{eqnarray}
    W_S^\text{gl,ELB} = 0.
\end{eqnarray}

In contrast, the LP \cite{Landi2021} approach keeps the weak coupling definition of internal energy (which for this model coincides with the minimal dissipation definition). Since internal energy is constant in time, change of internal energy will be always equal to zero \cite{Popovic2023}:
\begin{eqnarray} \label{eq:U_LP}
    \Delta U_S^\text{gl,LP} (t) = 0.
\end{eqnarray}
To reconcile this internal energy with the heat contribution in Eq. \eqref{eq:heat_ELB_example} and maintain the first law of thermodynamics, the work contribution has to compensate the heat contribution: 
\begin{eqnarray} \label{eq:work_LP}
    W_S^\text{gl,LP} (t) &=& 2 \int_0^\infty d\omega J(\omega) \frac{1-\cos(\omega t)}{\omega} \nonumber \\
    &=& 2 \alpha \Omega \frac{(\Omega t)^2}{1 + ( \Omega t)^2},
\end{eqnarray}
where in the second line we have substituted the spectral density in Eq. \eqref{eq:J_ohmic}.

\begin{figure}
    \centering
    \includegraphics[width=0.99\linewidth]{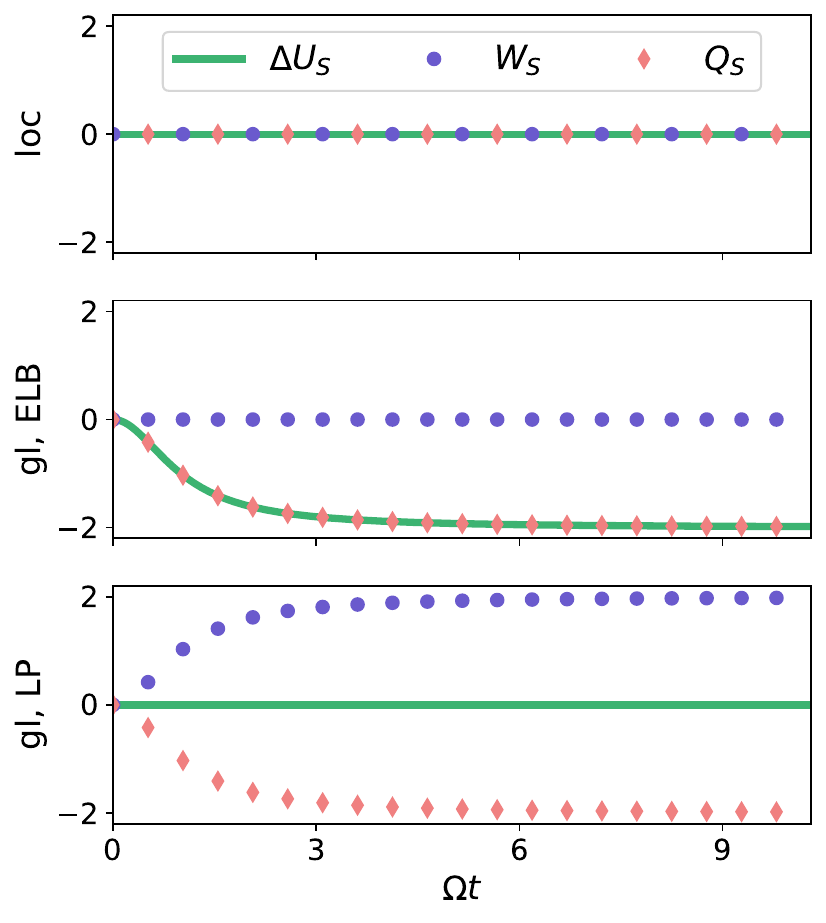}
    \caption{First law quantities, namely internal energy $U_S$, work $W_S$ and heat $Q_S$, as a function of time, for the local (a) and the global ELB (b) and LP (c) approaches, same parameters as in Fig. \ref{fig:second_law}.}
    \label{fig:first_law}
\end{figure}

With this we have written down all quantities pertaining the first law in the approaches studied here, and we are able to compare them. Starting with
heat: In the minimal dissipation case there is no heat exchange (Eq. \eqref{eq:heat_md_example}). According to this approach, the energy of the system in this model is stored in the populations, and since these do not change, energy remains constant. On the other hand, the global approach gives the heat contribution in Eq. \eqref{eq:heat_ELB_example}, which is time dependent and involves the spectral density of the environment. Interestingly, this quantity does not depend on $\omega_0$, the natural frequency and energy gap of the bare system, which clearly indicates that energy is exchanged only between the energy of the environment and the interaction energy. 
We plot these expressions and all other first-law quantities in Fig. \ref{fig:first_law}. 

Next, looking at internal energy, we see that minimal dissipation and the global LP approach both give in this model the same internal energy as the weak coupling formulation, which further makes internal energy constant in time. On the other hand, the global ELB approach amounts to considering the whole interaction Hamiltonian as part of the system, and hence the internal energy in Eq. \eqref{eq:U_example_ELB} gains an extra term and becomes time-dependent. 

In regard to work, now it is the global ELB and the local approach which coincide in that there is no work exchange between system and environment. Yet, this is a particular feature of this example model. In general, the minimal dissipation approach allows for work exchange between system and environment even when the microscopic Hamiltonian is time-independent \cite{Colla2022a, Colla2024Dec}, a feature not shared by the ELB approach. On the other hand, the global LP approach gives us a work contribution which exactly compensates the heat exchange. These quantities scale linearly with the frequency cutoff $\Omega$ and inverse temperature $\beta$, and can therefore become arbitrarily large.

\section{Discussion and conclusions}\label{sec:conclusions}

In this work we have studied pure decoherence scenarios, comparing local and global perspectives of quantum
thermodynamics, and we have seen that these can describe the same scenario very differently, providing 
different insights into the process. For example, from the local perspective entropy production is completely
determined by the change of von Neumann entropy and, thus, limited by the initial coherences of the open system. 
On the other hand, within the global approach entropy production is dominated by the contribution from
the heat flow and scales linearly with the cutoff frequency $\Omega$ of the spectral density. Correspondingly, 
while from the local point of view the energy of the system is constant in time and neither heat nor work is 
exchanged with the environment, from the global perspective one observes a large heat flow defined by the
change of energy of the bath, which is proportional to the cutoff frequency, indicating that essentially 
all reservoir oscillators commonly contribute to the heat flow. We note that this quantity refers to the 
rearranging of energy in the bath degrees of freedom, rather than to the energy entering or leaving the 
two-level system.

Finally, we mention an important fact which is nicely illustrated by the model discussed here. First, we note that the dephasing model introduced in 
Sec.~\ref{sec:example} has a 
well-defined Born-Markov limit for the open system dynamics. In fact, for times which are large compared to the thermal correlation time 
$\tau_E=\beta/\pi$, i.e. for $t \gg \tau_E$ the time dependent rate in the master equation \eqref{eq:master_equation_example} approaches the constant value \cite{Breuer2002}
\begin{eqnarray}
    \Gamma^M = \lim_{t\rightarrow \infty} \Gamma(t) = \frac{\alpha\pi}{\beta}
\end{eqnarray}
describing an exponential decay of the coherences of the open system's density matrix. The predictions of the local, minimal dissipation approach
expressed by Eqs.~\eqref{U_local}-\eqref{eq:heat_md_example} then obviously coincide with the standard weak coupling Markovian approach suggested by Spohn, Lebowitz and Alicki \cite{Spohn1978,Lebowitz1978, Alicki1979May}. However, within the global approach the thermodynamic quantities internal energy, work and heat are given by Eqs.~\eqref{eq:U_example_ELB}-\eqref{eq:work_LP} which obviously differ from the corresponding expressions for the weak coupling Markovian results. The same holds true for the entropy production as can be seen comparing 
Eqs.~\eqref{eq:ent_prod_md_example} and \eqref{entropy-production-global}. Thus, we conclude that even within second order in the system-environment coupling, i.e. to order $\alpha$, the local and the global approaches predict, in general, different expressions for work and heat and entropy production. It is not generally true, as one might conjecture, that the different approaches to quantum thermodynamics coincide in the limit of weak system-environment coupling.

Concluding, we remark that pure decoherence processes constitute a platform which is useful to compare different approaches to quantum thermodynamics.
Studies with more complex models and more complex dynamics should still be carried out in the future to better understand the difference between the approaches discussed here. Furthermore, there are many other general theoretical approaches in the literature (see, e.g., \cite{Elouard2023, Rivas2020}) for which a similar study as the one presented here should be performed.

\acknowledgments
The authors thank Tim Alh\"auser for many helpful discussions. I.A.P. acknowledges financial support from the Studienstiftung des deutschen Volkes, and
A.C. and H.P.B. acknowledge support from the European Union’s Framework Programme for Research and Innovation Horizon 2020 (2014-2020) 
under the Marie Skłodowska-Curie Grant Agreement No. 847471. A.C. also acknowledges support from MUR via the PRIN 2022 Project “Quantum Reservoir Computing
(QuReCo)” (contract n. 2022FEXLYB).

%%%%%%%%%%%%%%%%%%%%%%%%%%%%%%%%%%%%%%%%%%%%%%%%%%%%%%
\appendix

\section{Master equation for general dephasing models} \label{app:general_dephasing_ME}
We want to derive the general structure of the generator of a master equation for pure decoherence models. In order to do this, we first notice that any superoperator can be expanded into any orthonormal basis of the Hilbert space as written in Eq. \eqref{eq:general_generator}: 
\begin{eqnarray}
    \mathcal{L}_t \left[\rho_S \right] = \sum_{jkln} \gamma_{ijkn} |j \rangle \langle k | \rho_S |l\rangle \langle n |
\end{eqnarray}
(omitting the time dependence of the density matrix and the coefficients for a shorter notation). The Hermiticity preserving and trace annihilating properties of the generator result in the following restrictions on the coefficients: 
\begin{eqnarray}\label{eq:hermiticity_trace}
    \left( \mathcal{L}_t \left[X \right] \right)^\dagger = \mathcal{L}_t \left[X^\dagger \right] &\Rightarrow& \gamma_{jkln} = \gamma_{nlkj}^*, \\
    \Tr \{\mathcal{L}_t \left[X \right] \} = 0 &\Rightarrow &\sum_{j} \gamma_{jklj} = 0
\end{eqnarray}
In this case, we choose the basis of the system Hamiltonian $H_S$. We then apply the symmetry Eq. \eqref{eq:symmetry} and use the fact that the operator $S = \sum_i s_i |i \rangle \langle i |$ is diagonal in the same basis as $H_S$. Calculating both sides of Eq. \eqref{eq:symmetry} gives: 
\begin{eqnarray}
    \hspace{-0.5cm} U_\lambda \mathcal{L}_t \left[\rho_S \right] U_\lambda^\dagger &=& \sum_{jkln} \gamma_{jkln} e^{ i \lambda (s_j - s_n)} |j \rangle \langle k | \rho_S |l \rangle \langle n|,\\
    \hspace{-0.5cm}\mathcal{L}_t \left[ U_\lambda \rho_S U_\lambda^\dagger \right] &=& \sum_{jkln} \gamma_{jkln} e^{i\lambda (s_k - s_l)} |j \rangle \langle k |\rho_S |l \rangle \langle n|.
\end{eqnarray}
Since both expressions have to be equal for any value of $\lambda$, we can conclude that 
\begin{eqnarray}
    \gamma_{jkln} = \gamma_{jkln} \delta_{jk} \delta_{ln}.
\end{eqnarray}
Together with Eq. \eqref{eq:hermiticity_trace}, this results in the conditions found in 
Eq. \eqref{eq:conditions_e_coeffs} for the coefficients of the master equation, which allows us to write the master equation for any pure decoherence process in the form of Eq. \eqref{eq:generator_decoherence}.

\section{Dynamics of the system}\label{app:calculations_rho_s}

We start by writing the interaction Hamiltonian as 
\begin{equation}\label{eq:HI_AB}
    H_I = A \otimes B = \sigma_z \otimes \sum_k \left( g_k b_k^\dagger + g_k^* b_k \right),
\end{equation}
where $A = \sigma_z$ acts on $\mathcal{H}_S$ and $B = \sum_k \left( g_k b_k^\dagger + g_k^* b_k \right)$ acts on $\mathcal{H}_E$.  We then go to the interaction picture by applying the transformation $X \rightarrow \tilde{X} = e^{i H_0 t} X e^{-i H_0 t} $ to all operators, with $H_0 = H_S + H_E$. The interaction Hamiltonian thus transforms to 
\begin{eqnarray}\label{eq:HI_intpic}
    H_I \rightarrow \Tilde{H}_I(t)  = \Tilde{A} (t) \otimes \Tilde{B}(t)
\end{eqnarray}
with $\Tilde{A} (t) = \sigma_z$ time independent and
\begin{equation}
    \Tilde{B}(t) = \sum_k \left( g_k b_k^\dagger e^{i \omega_k t} + g_k^* b_k e^{-i \omega_k t} \right).
\end{equation}
Then, the von Neumann equation in interaction picture becomes $\frac{d}{dt} \tilde{\rho}_{SE} (t) = - i \left[ \tilde{H}_I(t) , \tilde{\rho}_{SE} (t) \right] $, and is solved through the unitary operator 
\begin{eqnarray} \label{eq:V_oft}
    V(t) &=& T_\leftarrow \exp \left[-i \int_0^t ds \tilde{H}_I (s) \right]  \\
    &=&  e^{i \phi(t) } \exp \left[ \sigma_z \otimes \sum_k \left( \alpha_k (t) b_k^\dagger - \alpha_k^* (t) b_k \right) \right], \nonumber
\end{eqnarray}
where we have defined $\alpha_k (t) \equiv g_k (1 - e^{i \omega_k t} )/ \omega_k$ and $\phi(t)$ is an overall, time-dependent phase factor \cite{Breuer2002}. To simplify the notation in the following, we specify the operator on the environment
\begin{eqnarray}
    \chi (t) \equiv \sum_k \left( \alpha_k (t) b_k^\dagger - \alpha_k^* (t) b_k \right).
\end{eqnarray}
The unitary operator \eqref{eq:V_oft} acts on both the system and the environment. To be able to work with it, we rewrite $\sigma_z$ in the energy eigenbasis of the bare system. Then, the unitary becomes
\begin{eqnarray} \label{eq:V_oft2}
    V(t) &=&  e^{i \phi(t) } e^{ \left(|1 \rangle \langle 1| - |0 \rangle \langle 0|\right) \otimes \chi (t) }\nonumber \\
     &=&e^{i \phi(t) } \underbrace{e^{|1 \rangle \langle 1| \otimes \chi (t)}}_{\equiv V_1(t)} \underbrace{e^{-|0 \rangle \langle 0| \otimes \chi(t)}}_{\equiv V_0(t)}  
\end{eqnarray}

We can rewrite the first of these terms by expanding the exponential: 
\begin{eqnarray}
    V_1(t) &=& e^{|1 \rangle \langle 1| \otimes \chi(t) } \nonumber \\
    &=& \sum_{n=0}^\infty \frac{1}{n!} (|1 \rangle \langle 1|)^n \otimes (\chi(t))^n \nonumber \\
    &=& \mathbbm{1} + \sum_{n=1}^\infty \frac{1}{n!} |1 \rangle \langle 1| \otimes (\chi(t))^n \nonumber \\
    &=& \mathbbm{1} + |1 \rangle \langle 1| \otimes \sum_{n=0}^\infty \frac{1}{n!}  (\chi(t))^n -  |1 \rangle \langle 1| \nonumber \\
    &=& |0 \rangle \langle 0| + |1 \rangle \langle 1| \otimes e^{ \chi(t)}.   
\end{eqnarray}
Following the same procedure we obtain 
\begin{eqnarray}
    V_0(t) = |1 \rangle \langle 1| + |0 \rangle \langle 0| \otimes e^{ -\chi(t)} .
\end{eqnarray}
With these expressions we see the action of the unitary operator on the system and environmental sides separately. Multiplying $V_0$ and $V_1$, the unitary governing the evolution reads
\begin{eqnarray}\label{eq:V_oft_separate}
    \hspace{-0.5cm}V(t) &=&  e^{i \phi(t) } \left(|1 \rangle \langle 1| \otimes e^{ \chi(t) } + |0 \rangle \langle 0| \otimes e^{- \chi(t) } \right).
\end{eqnarray}

The dynamics of the density matrix in interaction picture is therefore given by 
\begin{equation}\label{eq:rho_SE_tilde}
    \tilde{\rho}_{SE} (t) = V(t) \left( \rho_S(0) \otimes \rho_E(0) \right) V^\dagger (t).
\end{equation}
Plugging our obtained expression for the unitary \eqref{eq:V_oft_separate} into \eqref{eq:rho_SE_tilde} we find 
\begin{eqnarray}\label{eq:rho_SE_tilde_explicit}
    \tilde{\rho}_{SE} &=& \rho_S^{00} (0) |0\rangle \langle 0 | e^{-\chi(t)} \rho_E(0) e^{\chi(t) } \nonumber \\
    &&+ \rho_S^{01} (0) |0 \rangle \langle 1 | e^{- \chi(t) } \rho_E(0) e^{- \chi(t)} \nonumber \\
    &&+ \rho_S^{10}(0)  |1 \rangle \langle 0 | e^{\chi(t) } \rho_E(0) e^{\chi(t)} \nonumber \\
    &&+ \rho_S^{11} (0)|1 \rangle \langle 1 | e^{\chi(t) } \rho_E (0) e^{-\chi(t)},
\end{eqnarray}
where the superindices indicate the matrix elements $\rho_S^{ij} = \langle i | \rho_S | j\rangle$. Next, we transform back to the Schrödinger picture and take the partial trace over the environment to get the reduced system density matrix at each point in time: 
\begin{eqnarray}\label{eq:rho_S_oft}
    \rho_{S}(t) = \Tr_E \{ e^{-i H_0 t} \tilde{\rho}_{SE} (t) e^{i H_0 t} \}. 
\end{eqnarray}
The cyclic property of the trace will take care of the environmental part of $H_0$, as well as of the exponentials of $\chi(t) $ for the diagonal elements (see Eq. \eqref{eq:rho_SE_tilde_explicit}). The reduced density matrix of the system thus takes the form 
\begin{eqnarray}
    \rho_S(t) &=& \rho_S^{00}(0) e^{-i H_S t} |0\rangle \langle 0 | e^{i H_S t} \nonumber \\
    &&+  \rho_S^{01}(0) e^{-i H_S t} |0\rangle \langle 1 | e^{i H_S t} \; \Tr_E \{e^{-2 \chi(t)} \rho_E(0) \}  \nonumber \\
    &&+  \rho_S^{10}(0) e^{-i H_S t} |1\rangle \langle 0| e^{i H_S t} \; \Tr_E \{e^{2 \chi(t)} \rho_E(0) \}  \nonumber \\
    &&+  \rho_S^{11}(0) e^{-i H_S t} |1\rangle \langle 1 | e^{i H_S t} .
\end{eqnarray}
Using $e^{i H_S t} = \cos \left( \frac{\omega_0}{2} t\right) \mathbbm{1} + i \sin \left(\frac{\omega_0}{2} t\right) \sigma_z$, the former equation can be written as 
\begin{eqnarray}\label{eq:rho_S_oft2}
    \rho_S(t) &=& \rho_S^{00}(0) |0\rangle \langle 0 | + \rho_S^{11}(0) |1\rangle \langle 1 | \nonumber \\
    &&+  \rho_S^{01}(0) e^{i \omega_0 t} |0\rangle \langle 1 | \; F_{-}(t)  \nonumber \\
    &&+  \rho_S^{10}(0) e^{-i \omega_0 t}  |1\rangle \langle 0|  \; F_+ (t),
\end{eqnarray}
using the function 
\begin{eqnarray}
    F_{\pm} (t) \equiv \Tr_E \left\{ e^{\pm 2 \chi(t) } \rho_E(0) \right\}.
\end{eqnarray}
From expression \eqref{eq:rho_S_oft2} we can already see that the populations will stay constant in time. To find the time dependence of the coherences, we need to notice that the functions $F_\pm (t)$ are the expectation value of time-dependent displacement operators over the initial thermal environment: 
\begin{eqnarray}
    F_{\pm} (t) &=& \Tr_E \left\{ e^{\sum_k \pm 2 \left( \alpha_k (t) b_k^\dagger - \alpha_k^* (t) b_k \right)\rho_E(0)} \right\}  \nonumber \\
    &=& \Tr_E \left\{ \bigotimes_k D_k(\pm 2 \alpha_k(t)) \rho_E(0) \right\},
\end{eqnarray}
where $D_k(\alpha) = \exp \left( \alpha b_k^\dagger - \alpha b_k \right)$ is the displacement operator acting on mode $k$ with a displacement $\alpha$. This is the Wigner characteristic function. Since the environment is initially in a Gaussian state, we find \cite{Serafini2017Jul}:  
\begin{eqnarray}
    &&F_\pm (t) = \bigotimes_k \exp \left[ - |\alpha_k(t)|^2 \coth \left(\frac{\beta \omega_k}{2}\right) \right] \\
    &&= \exp \left[ - \sum_k \frac{2|g_k|^2}{\omega_k^2} (1 - \cos (\omega_k t))\coth \left(\frac{\beta \omega_k}{2}\right) \right] . \nonumber
\end{eqnarray}
Using the definition of the spectral density in Eq. \eqref{eq:J_def} and introducing the decoherence function $\eta(t) $ such that $F_\pm(t) \equiv e^{-\eta(t)}$ we obtain 
\begin{eqnarray}
    \eta(t) = 2 \int_0^\infty d\omega J(\omega) \frac{1-\cos(\omega t)}{\omega^2} \coth\left(\frac{\beta \omega}{2}\right).
\end{eqnarray}

\section{Derivation of the master equation}
\label{app:master_equation}
We show how to derive the master equation for the dynamics of the system using the exact solution \eqref{eq:rho_S_00}-\eqref{eq:rho_S_10}. This method can be generally applied when an exact solution is known \cite{Colla2024Dec, Picatoste2024Mar}, and the idea behind it is to compare the exact solution to an ansatz master equation with some unknown coefficients, and match the results to obtain expressions for the coefficients. 

We start by considering the time derivative of the expectation values of a set of system operators forming an operator basis,
which can be obtained from the general formula
\begin{eqnarray}
    \frac{d}{dt} \langle O_S \rangle_t = \Tr \{ O_S \dot{\rho}_S(t) \},
\end{eqnarray}
yielding
\begin{eqnarray}
    \frac{d}{dt} \langle \sigma_z \rangle_t &=& 0 ,\label{eq:sigma_z_dot_solution}\\
    \frac{d}{dt} \langle \sigma_+ \rangle_t &=& \left[ i \omega_0 - \dot{\eta} (t) \right] \langle \sigma_+ \rangle_t,\label{eq:sigma_p_dot_solution}
\end{eqnarray}
where we recall that $\dot{\eta}(t) $ is a real function. 

Alongside, we have to find an ansatz for the structure of the master equation. Considering the dephasing dynamics of the system, we propose the following: 
\begin{eqnarray}
    \frac{d}{dt} \rho_S(t) &=& -i \left[ \Omega (t) \sigma_z , \rho_S(t) \right] \nonumber\\
    &&+ \Gamma(t) \left[ \sigma_z \rho_S(t) \sigma_z - \rho_S(t) \right],
\end{eqnarray}
where $\Omega(t)$ and $\Gamma(t)$ have to be real functions due to the Hermiticity of the density matrix. We calculate the time derivative of the expectation values of the Pauli matrices from the ansatz master equation and obtain 
\begin{eqnarray}
    \frac{d}{dt} \langle \sigma_z \rangle_t &=& 0 ,\label{eq:sigma_z_dot_ansatz}\\
    \frac{d}{dt} \langle \sigma_+ \rangle_t &=& \left[ 2 i \Omega(t) - 2 \Gamma (t) \right] \langle \sigma_+ \rangle_t  . \label{eq:sigma_p_dot_ansatz}
\end{eqnarray}
The strategy is now to determine the coefficients $\Omega (t), \Gamma(t)$ by comparing equations \eqref{eq:sigma_z_dot_ansatz}- \eqref{eq:sigma_p_dot_ansatz} with the ones obtained from the exact solution \eqref{eq:sigma_z_dot_solution}-\eqref{eq:sigma_p_dot_solution}. This gives the following: 
\begin{eqnarray}
    \Omega(t) &=& \omega_0 / 2, \\
    \Gamma(t) &=& \frac{\dot{\eta} (t)}{2} 
\end{eqnarray}
Thus, we can write down the exact master equation for the dynamics of the system.

Another strategy, specific to this kind of model and environment, would be to use the TCL expansion \cite{Breuer2002}, since it was proven that, for dephasing interaction on the system side and an initially zero-mean Gaussian environment, the second order TCL master equation is exact \cite{Ban2010, Doll2008, Wissmann2016}. It can be shown that the same result is obtained in that case. 

\section{Evolution of the expectation value of the interaction Hamiltonian}\label{app:calculations_HI}

In order to calculate the expectation value of the interaction Hamiltonian in time, we use the exact solution found in Appendix \ref{app:calculations_rho_s}. With the expression found in \eqref{eq:rho_SE_tilde}, and transforming the density matrix back to Schrödinger picture, we have
\begin{eqnarray}
    \langle H_I \rangle_t &=& \Tr \left\{H_I \rho_{SE}(t)\right\} \nonumber\\
    &=& \Tr \left\{ H_I e^{-i H_0 t} V(t) \rho_{SE}(0) V^\dagger (t) e^{i H_0 t} \right\} \nonumber\\
    &=& \Tr \left\{V^\dagger (t) e^{i H_0 t}  H_I e^{-i H_0 t} V(t) \rho_{SE}(0) \right\} \nonumber \\
    &=& \Tr \left\{V^\dagger (t) \tilde{H}_I(t) V(t) \rho_{SE}(0) \right\},
\end{eqnarray}
where we have used the interaction Hamiltonian in interaction picture defined in Eq. \eqref{eq:HI_intpic}. Then, plugging in the expression for the unitary operator found in Eq. \eqref{eq:V_oft_separate} we find 
\begin{eqnarray}
    \langle H_I \rangle_t &=& \rho_S^{11}(0) \Tr_E \{e^{- \chi(t)} \tilde{B}(t) e^{\chi(t)} \rho_E(0) \} \nonumber \\
    &&-\rho_S^{00}(0) \Tr_E \{e^{ \chi(t)} \tilde{B}(t) e^{-\chi(t)} \rho_E(0) \} .
\end{eqnarray}
Again we make use of the fact that $e^{\chi(t)}$ is a collective displacement operator, such that $e^{\chi(t) } = \bigotimes_k D_k (\alpha_k(t))$, we see that 
\begin{eqnarray}
    e^{\chi(t)} b_k e^{- \chi(t)} &=& \bigotimes_j D_j (\alpha_j(t)) b_k \bigotimes_l D_l^\dagger (\alpha_l(t)) 
    \nonumber \\
    &=& b_k + \alpha_k(t).
\end{eqnarray}
Since the intial state is thermal, meaning $\Tr \{b_k \rho_E(0) \} = 0$, we get the final result
\begin{eqnarray}
    \langle H_I \rangle_t &=& (\rho_S^{11}(0) + \rho_S^{00}(0)) \sum_k g_k \alpha_k^* e^{i \omega_k t} + g_k^* \alpha_k e^{-i \omega_k t} \nonumber \\
    &=& -2 \int_0^t d \omega J(\omega) \frac{1- \cos(\omega t)}{\omega},
\end{eqnarray}
where we have replaced the sum over the modes by the integral over the spectral density using Eq. \eqref{eq:J_def}.

%%%%%%%%%%%%%%%%%%%%%%%%%%%%%%%%%%%%%%%%%%%%%%%%%%%%%%%%%%%%%%%%%

\bibliography{biblio}

\begin{thebibliography}{40}%
\makeatletter
\providecommand \@ifxundefined [1]{%
 \@ifx{#1\undefined}
}%
\providecommand \@ifnum [1]{%
 \ifnum #1\expandafter \@firstoftwo
 \else \expandafter \@secondoftwo
 \fi
}%
\providecommand \@ifx [1]{%
 \ifx #1\expandafter \@firstoftwo
 \else \expandafter \@secondoftwo
 \fi
}%
\providecommand \natexlab [1]{#1}%
\providecommand \enquote  [1]{``#1''}%
\providecommand \bibnamefont  [1]{#1}%
\providecommand \bibfnamefont [1]{#1}%
\providecommand \citenamefont [1]{#1}%
\providecommand \href@noop [0]{\@secondoftwo}%
\providecommand \href [0]{\begingroup \@sanitize@url \@href}%
\providecommand \@href[1]{\@@startlink{#1}\@@href}%
\providecommand \@@href[1]{\endgroup#1\@@endlink}%
\providecommand \@sanitize@url [0]{\catcode `\\12\catcode `\$12\catcode `\&12\catcode `\#12\catcode `\^12\catcode `\_12\catcode `\%12\relax}%
\providecommand \@@startlink[1]{}%
\providecommand \@@endlink[0]{}%
\providecommand \url  [0]{\begingroup\@sanitize@url \@url }%
\providecommand \@url [1]{\endgroup\@href {#1}{\urlprefix }}%
\providecommand \urlprefix  [0]{URL }%
\providecommand \Eprint [0]{\href }%
\providecommand \doibase [0]{http://dx.doi.org/}%
\providecommand \selectlanguage [0]{\@gobble}%
\providecommand \bibinfo  [0]{\@secondoftwo}%
\providecommand \bibfield  [0]{\@secondoftwo}%
\providecommand \translation [1]{[#1]}%
\providecommand \BibitemOpen [0]{}%
\providecommand \bibitemStop [0]{}%
\providecommand \bibitemNoStop [0]{.\EOS\space}%
\providecommand \EOS [0]{\spacefactor3000\relax}%
\providecommand \BibitemShut  [1]{\csname bibitem#1\endcsname}%
\let\auto@bib@innerbib\@empty
%</preamble>
\bibitem [{\citenamefont {Gemmer}\ \emph {et~al.}(2004)\citenamefont {Gemmer}, \citenamefont {Michel},\ and\ \citenamefont {Mahler}}]{Gemmer2004}%
  \BibitemOpen
  \bibfield  {author} {\bibinfo {author} {\bibfnamefont {J.}~\bibnamefont {Gemmer}}, \bibinfo {author} {\bibfnamefont {M.}~\bibnamefont {Michel}}, \ and\ \bibinfo {author} {\bibfnamefont {G.}~\bibnamefont {Mahler}},\ }\href@noop {} {\emph {\bibinfo {title} {{Quantum Thermodynamics}}}}\ (\bibinfo  {publisher} {Springer},\ \bibinfo {address} {Berlin},\ \bibinfo {year} {2004})\BibitemShut {NoStop}%
\bibitem [{\citenamefont {Binder}\ \emph {et~al.}(2018)\citenamefont {Binder}, \citenamefont {Correa}, \citenamefont {Gogolin}, \citenamefont {Anders},\ and\ \citenamefont {Adesso}}]{Binder2018}%
  \BibitemOpen
  \bibfield  {author} {\bibinfo {author} {\bibfnamefont {F.}~\bibnamefont {Binder}}, \bibinfo {author} {\bibfnamefont {L.~A.}\ \bibnamefont {Correa}}, \bibinfo {author} {\bibfnamefont {C.}~\bibnamefont {Gogolin}}, \bibinfo {author} {\bibfnamefont {J.}~\bibnamefont {Anders}}, \ and\ \bibinfo {author} {\bibfnamefont {G.}~\bibnamefont {Adesso}},\ }\href@noop {} {\emph {\bibinfo {title} {Thermodynamics in the Quantum Regime}}}\ (\bibinfo  {publisher} {Springer},\ \bibinfo {address} {Cham, Switzerland},\ \bibinfo {year} {2018})\BibitemShut {NoStop}%
\bibitem [{\citenamefont {Breuer}\ and\ \citenamefont {Petruccione}(2002)}]{Breuer2002}%
  \BibitemOpen
  \bibfield  {author} {\bibinfo {author} {\bibfnamefont {H.-P.}\ \bibnamefont {Breuer}}\ and\ \bibinfo {author} {\bibfnamefont {F.}~\bibnamefont {Petruccione}},\ }\href@noop {} {\emph {\bibinfo {title} {The Theory of Open Quantum Systems}}}\ (\bibinfo  {publisher} {Oxford University Press},\ \bibinfo {address} {Oxford},\ \bibinfo {year} {2002})\BibitemShut {NoStop}%
\bibitem [{\citenamefont {Esposito}\ \emph {et~al.}(2010)\citenamefont {Esposito}, \citenamefont {Lindenberg},\ and\ \citenamefont {den Broeck}}]{Esposito2010}%
  \BibitemOpen
  \bibfield  {author} {\bibinfo {author} {\bibfnamefont {M.}~\bibnamefont {Esposito}}, \bibinfo {author} {\bibfnamefont {K.}~\bibnamefont {Lindenberg}}, \ and\ \bibinfo {author} {\bibfnamefont {C.~V.}\ \bibnamefont {den Broeck}},\ }\href {\doibase 10.1088/1367-2630/12/1/013013} {\bibfield  {journal} {\bibinfo  {journal} {New Journal of Physics}\ }\textbf {\bibinfo {volume} {12}},\ \bibinfo {pages} {013013} (\bibinfo {year} {2010})}\BibitemShut {NoStop}%
\bibitem [{\citenamefont {Colla}(2024)}]{CollaPhD}%
  \BibitemOpen
  \bibfield  {author} {\bibinfo {author} {\bibfnamefont {A.}~\bibnamefont {Colla}},\ }\href@noop {} {\enquote {\bibinfo {title} {Dynamically emergent quantum thermodynamics of open systems},}\ }\bibinfo {howpublished} {\url{https://freidok.uni-freiburg.de/data/263695}} (\bibinfo {year} {2024})\BibitemShut {NoStop}%
\bibitem [{\citenamefont {Strasberg}\ \emph {et~al.}(2021)\citenamefont {Strasberg}, \citenamefont {D\'{\i}az},\ and\ \citenamefont {Riera-Campeny}}]{Strasberg2021}%
  \BibitemOpen
  \bibfield  {author} {\bibinfo {author} {\bibfnamefont {P.}~\bibnamefont {Strasberg}}, \bibinfo {author} {\bibfnamefont {M.~G.}\ \bibnamefont {D\'{\i}az}}, \ and\ \bibinfo {author} {\bibfnamefont {A.}~\bibnamefont {Riera-Campeny}},\ }\href {\doibase 10.1103/PhysRevE.104.L022103} {\bibfield  {journal} {\bibinfo  {journal} {Phys. Rev. E}\ }\textbf {\bibinfo {volume} {104}},\ \bibinfo {pages} {L022103} (\bibinfo {year} {2021})}\BibitemShut {NoStop}%
\bibitem [{\citenamefont {Deffner}\ and\ \citenamefont {Lutz}(2010)}]{Deffner2010}%
  \BibitemOpen
  \bibfield  {author} {\bibinfo {author} {\bibfnamefont {S.}~\bibnamefont {Deffner}}\ and\ \bibinfo {author} {\bibfnamefont {E.}~\bibnamefont {Lutz}},\ }\href {\doibase 10.1103/PhysRevLett.105.170402} {\bibfield  {journal} {\bibinfo  {journal} {Phys. Rev. Lett.}\ }\textbf {\bibinfo {volume} {105}},\ \bibinfo {pages} {170402} (\bibinfo {year} {2010})}\BibitemShut {NoStop}%
\bibitem [{\citenamefont {Nicacio}\ and\ \citenamefont {Maia}(2023)}]{Nicacio2023}%
  \BibitemOpen
  \bibfield  {author} {\bibinfo {author} {\bibfnamefont {F.}~\bibnamefont {Nicacio}}\ and\ \bibinfo {author} {\bibfnamefont {R.~N.~P.}\ \bibnamefont {Maia}},\ }\href {\doibase 10.1103/PhysRevA.108.022209} {\bibfield  {journal} {\bibinfo  {journal} {Phys. Rev. A}\ }\textbf {\bibinfo {volume} {108}},\ \bibinfo {pages} {022209} (\bibinfo {year} {2023})}\BibitemShut {NoStop}%
\bibitem [{\citenamefont {Landi}\ and\ \citenamefont {Paternostro}(2021)}]{Landi2021}%
  \BibitemOpen
  \bibfield  {author} {\bibinfo {author} {\bibfnamefont {G.~T.}\ \bibnamefont {Landi}}\ and\ \bibinfo {author} {\bibfnamefont {M.}~\bibnamefont {Paternostro}},\ }\href {\doibase 10.1103/RevModPhys.93.035008} {\bibfield  {journal} {\bibinfo  {journal} {Rev. Mod. Phys.}\ }\textbf {\bibinfo {volume} {93}},\ \bibinfo {pages} {035008} (\bibinfo {year} {2021})}\BibitemShut {NoStop}%
\bibitem [{\citenamefont {Rivas}(2020)}]{Rivas2020}%
  \BibitemOpen
  \bibfield  {author} {\bibinfo {author} {\bibfnamefont {A.}~\bibnamefont {Rivas}},\ }\href {\doibase 10.1103/PhysRevLett.124.160601} {\bibfield  {journal} {\bibinfo  {journal} {Phys. Rev. Lett.}\ }\textbf {\bibinfo {volume} {124}},\ \bibinfo {pages} {160601} (\bibinfo {year} {2020})}\BibitemShut {NoStop}%
\bibitem [{\citenamefont {Elouard}\ and\ \citenamefont {Lombard~Latune}(2023)}]{Elouard2023}%
  \BibitemOpen
  \bibfield  {author} {\bibinfo {author} {\bibfnamefont {C.}~\bibnamefont {Elouard}}\ and\ \bibinfo {author} {\bibfnamefont {C.}~\bibnamefont {Lombard~Latune}},\ }\href {\doibase 10.1103/PRXQuantum.4.020309} {\bibfield  {journal} {\bibinfo  {journal} {PRX Quantum}\ }\textbf {\bibinfo {volume} {4}},\ \bibinfo {pages} {020309} (\bibinfo {year} {2023})}\BibitemShut {NoStop}%
\bibitem [{\citenamefont {Alicki}\ and\ \citenamefont {Lendi}(1987)}]{Alicki1987}%
  \BibitemOpen
  \bibfield  {author} {\bibinfo {author} {\bibfnamefont {R.}~\bibnamefont {Alicki}}\ and\ \bibinfo {author} {\bibfnamefont {K.}~\bibnamefont {Lendi}},\ }\href@noop {} {\emph {\bibinfo {title} {Quantum Dynamical Semigroups and Applications}}},\ \bibinfo {series} {Lecture Notes in Physics}, Vol.\ \bibinfo {volume} {286}\ (\bibinfo  {publisher} {Springer},\ \bibinfo {address} {Berlin},\ \bibinfo {year} {1987})\BibitemShut {NoStop}%
\bibitem [{\citenamefont {Colla}\ and\ \citenamefont {Breuer}(2022)}]{Colla2022a}%
  \BibitemOpen
  \bibfield  {author} {\bibinfo {author} {\bibfnamefont {A.}~\bibnamefont {Colla}}\ and\ \bibinfo {author} {\bibfnamefont {H.-P.}\ \bibnamefont {Breuer}},\ }\href {\doibase 10.1103/PhysRevA.105.052216} {\bibfield  {journal} {\bibinfo  {journal} {Phys. Rev. A}\ }\textbf {\bibinfo {volume} {105}},\ \bibinfo {pages} {052216} (\bibinfo {year} {2022})}\BibitemShut {NoStop}%
\bibitem [{\citenamefont {Spohn}(1978)}]{Spohn1978}%
  \BibitemOpen
  \bibfield  {author} {\bibinfo {author} {\bibfnamefont {H.}~\bibnamefont {Spohn}},\ }\href {\doibase 10.1063/1.523789} {\bibfield  {journal} {\bibinfo  {journal} {Journal of Mathematical Physics}\ }\textbf {\bibinfo {volume} {19}},\ \bibinfo {pages} {1227} (\bibinfo {year} {1978})},\ \Eprint {http://arxiv.org/abs/https://doi.org/10.1063/1.523789} {https://doi.org/10.1063/1.523789} \BibitemShut {NoStop}%
\bibitem [{\citenamefont {Popovic}\ \emph {et~al.}(2023)\citenamefont {Popovic}, \citenamefont {Mitchison},\ and\ \citenamefont {Goold}}]{Popovic2023}%
  \BibitemOpen
  \bibfield  {author} {\bibinfo {author} {\bibfnamefont {M.}~\bibnamefont {Popovic}}, \bibinfo {author} {\bibfnamefont {M.~T.}\ \bibnamefont {Mitchison}}, \ and\ \bibinfo {author} {\bibfnamefont {J.}~\bibnamefont {Goold}},\ }\href {\doibase 10.1098/rspa.2023.0040} {\bibfield  {journal} {\bibinfo  {journal} {Proc. R. Soc. A.}\ }\textbf {\bibinfo {volume} {479}} (\bibinfo {year} {2023}),\ 10.1098/rspa.2023.0040}\BibitemShut {NoStop}%
\bibitem [{\citenamefont {Marcantoni}(2017)}]{Marcantoni2017-tl}%
  \BibitemOpen
  \bibfield  {author} {\bibinfo {author} {\bibfnamefont {S.}~\bibnamefont {Marcantoni}},\ }\href@noop {} {\bibfield  {journal} {\bibinfo  {journal} {J. Phys. Conf. Ser.}\ }\textbf {\bibinfo {volume} {841}},\ \bibinfo {pages} {012019} (\bibinfo {year} {2017})}\BibitemShut {NoStop}%
\bibitem [{\citenamefont {Colla}\ \emph {et~al.}(2022)\citenamefont {Colla}, \citenamefont {Neubrand},\ and\ \citenamefont {Breuer}}]{Colla2022b}%
  \BibitemOpen
  \bibfield  {author} {\bibinfo {author} {\bibfnamefont {A.}~\bibnamefont {Colla}}, \bibinfo {author} {\bibfnamefont {N.}~\bibnamefont {Neubrand}}, \ and\ \bibinfo {author} {\bibfnamefont {H.-P.}\ \bibnamefont {Breuer}},\ }\href {\doibase 10.1088/1367-2630/aca709} {\bibfield  {journal} {\bibinfo  {journal} {New Journal of Physics}\ }\textbf {\bibinfo {volume} {24}},\ \bibinfo {pages} {123005} (\bibinfo {year} {2022})}\BibitemShut {NoStop}%
\bibitem [{\citenamefont {Vacchini}\ and\ \citenamefont {Amato}(2016)}]{Vacchini2016-se}%
  \BibitemOpen
  \bibfield  {author} {\bibinfo {author} {\bibfnamefont {B.}~\bibnamefont {Vacchini}}\ and\ \bibinfo {author} {\bibfnamefont {G.}~\bibnamefont {Amato}},\ }\href@noop {} {\bibfield  {journal} {\bibinfo  {journal} {Sci. Rep.}\ }\textbf {\bibinfo {volume} {6}} (\bibinfo {year} {2016})}\BibitemShut {NoStop}%
\bibitem [{\citenamefont {Alipour}\ \emph {et~al.}(2020)\citenamefont {Alipour}, \citenamefont {Rezakhani}, \citenamefont {Babu}, \citenamefont {M\o{}lmer}, \citenamefont {M\"ott\"onen},\ and\ \citenamefont {Ala-Nissila}}]{Alipour2020}%
  \BibitemOpen
  \bibfield  {author} {\bibinfo {author} {\bibfnamefont {S.}~\bibnamefont {Alipour}}, \bibinfo {author} {\bibfnamefont {A.~T.}\ \bibnamefont {Rezakhani}}, \bibinfo {author} {\bibfnamefont {A.~P.}\ \bibnamefont {Babu}}, \bibinfo {author} {\bibfnamefont {K.}~\bibnamefont {M\o{}lmer}}, \bibinfo {author} {\bibfnamefont {M.}~\bibnamefont {M\"ott\"onen}}, \ and\ \bibinfo {author} {\bibfnamefont {T.}~\bibnamefont {Ala-Nissila}},\ }\href {\doibase 10.1103/PhysRevX.10.041024} {\bibfield  {journal} {\bibinfo  {journal} {Phys. Rev. X}\ }\textbf {\bibinfo {volume} {10}},\ \bibinfo {pages} {041024} (\bibinfo {year} {2020})}\BibitemShut {NoStop}%
\bibitem [{\citenamefont {Shibata}\ \emph {et~al.}(1977)\citenamefont {Shibata}, \citenamefont {Takahashi},\ and\ \citenamefont {Hashitsume}}]{Shibata1977}%
  \BibitemOpen
  \bibfield  {author} {\bibinfo {author} {\bibfnamefont {F.}~\bibnamefont {Shibata}}, \bibinfo {author} {\bibfnamefont {Y.}~\bibnamefont {Takahashi}}, \ and\ \bibinfo {author} {\bibfnamefont {N.}~\bibnamefont {Hashitsume}},\ }\href@noop {} {\bibfield  {journal} {\bibinfo  {journal} {J. Stat. Phys.}\ }\textbf {\bibinfo {volume} {17}},\ \bibinfo {pages} {171} (\bibinfo {year} {1977})}\BibitemShut {NoStop}%
\bibitem [{\citenamefont {Chaturvedi}\ and\ \citenamefont {Shibata}(1979)}]{Shibata1979}%
  \BibitemOpen
  \bibfield  {author} {\bibinfo {author} {\bibfnamefont {S.}~\bibnamefont {Chaturvedi}}\ and\ \bibinfo {author} {\bibfnamefont {F.}~\bibnamefont {Shibata}},\ }\href@noop {} {\bibfield  {journal} {\bibinfo  {journal} {Z. Phys. B}\ }\textbf {\bibinfo {volume} {35}},\ \bibinfo {pages} {297} (\bibinfo {year} {1979})}\BibitemShut {NoStop}%
\bibitem [{\citenamefont {Hall}\ \emph {et~al.}(2014)\citenamefont {Hall}, \citenamefont {Cresser}, \citenamefont {Li},\ and\ \citenamefont {Andersson}}]{Hall2014Apr}%
  \BibitemOpen
  \bibfield  {author} {\bibinfo {author} {\bibfnamefont {M.~J.~W.}\ \bibnamefont {Hall}}, \bibinfo {author} {\bibfnamefont {J.~D.}\ \bibnamefont {Cresser}}, \bibinfo {author} {\bibfnamefont {L.}~\bibnamefont {Li}}, \ and\ \bibinfo {author} {\bibfnamefont {E.}~\bibnamefont {Andersson}},\ }\href {\doibase 10.1103/PhysRevA.89.042120} {\bibfield  {journal} {\bibinfo  {journal} {Phys. Rev. A}\ }\textbf {\bibinfo {volume} {89}},\ \bibinfo {pages} {042120} (\bibinfo {year} {2014})}\BibitemShut {NoStop}%
\bibitem [{\citenamefont {Breuer}\ \emph {et~al.}(1999)\citenamefont {Breuer}, \citenamefont {Kappler},\ and\ \citenamefont {Petruccione}}]{Breuer1999b}%
  \BibitemOpen
  \bibfield  {author} {\bibinfo {author} {\bibfnamefont {H.-P.}\ \bibnamefont {Breuer}}, \bibinfo {author} {\bibfnamefont {B.}~\bibnamefont {Kappler}}, \ and\ \bibinfo {author} {\bibfnamefont {F.}~\bibnamefont {Petruccione}},\ }\href@noop {} {\bibfield  {journal} {\bibinfo  {journal} {Phys. Rev. A}\ }\textbf {\bibinfo {volume} {59}},\ \bibinfo {pages} {1633} (\bibinfo {year} {1999})}\BibitemShut {NoStop}%
\bibitem [{\citenamefont {Breuer}\ \emph {et~al.}(2016)\citenamefont {Breuer}, \citenamefont {Laine}, \citenamefont {Piilo},\ and\ \citenamefont {Vacchini}}]{Breuer2016a}%
  \BibitemOpen
  \bibfield  {author} {\bibinfo {author} {\bibfnamefont {H.-P.}\ \bibnamefont {Breuer}}, \bibinfo {author} {\bibfnamefont {E.-M.}\ \bibnamefont {Laine}}, \bibinfo {author} {\bibfnamefont {J.}~\bibnamefont {Piilo}}, \ and\ \bibinfo {author} {\bibfnamefont {B.}~\bibnamefont {Vacchini}},\ }\href@noop {} {\bibfield  {journal} {\bibinfo  {journal} {Rev. Mod. Phys.}\ }\textbf {\bibinfo {volume} {88}},\ \bibinfo {pages} {021002} (\bibinfo {year} {2016})}\BibitemShut {NoStop}%
\bibitem [{\citenamefont {Strasberg}\ and\ \citenamefont {Esposito}(2019)}]{Strasberg2019}%
  \BibitemOpen
  \bibfield  {author} {\bibinfo {author} {\bibfnamefont {P.}~\bibnamefont {Strasberg}}\ and\ \bibinfo {author} {\bibfnamefont {M.}~\bibnamefont {Esposito}},\ }\href {\doibase 10.1103/PhysRevE.99.012120} {\bibfield  {journal} {\bibinfo  {journal} {Phys. Rev. E}\ }\textbf {\bibinfo {volume} {99}},\ \bibinfo {pages} {012120} (\bibinfo {year} {2019})}\BibitemShut {NoStop}%
\bibitem [{\citenamefont {Parthasarathy}(1992)}]{Parthasarathy1992}%
  \BibitemOpen
  \bibfield  {author} {\bibinfo {author} {\bibfnamefont {K.~R.}\ \bibnamefont {Parthasarathy}},\ }\href {https://link.springer.com/book/10.1007/978-3-0348-0566-7} {\emph {\bibinfo {title} {{An Introduction to Quantum Stochastic Calculus}}}}\ (\bibinfo  {publisher} {Springer},\ \bibinfo {address} {Basel, Switzerland},\ \bibinfo {year} {1992})\BibitemShut {NoStop}%
\bibitem [{\citenamefont {Hayden}\ and\ \citenamefont {Sorce}(2022)}]{Hayden2022May}%
  \BibitemOpen
  \bibfield  {author} {\bibinfo {author} {\bibfnamefont {P.}~\bibnamefont {Hayden}}\ and\ \bibinfo {author} {\bibfnamefont {J.}~\bibnamefont {Sorce}},\ }\href {\doibase 10.1088/1751-8121/ac65c2} {\bibfield  {journal} {\bibinfo  {journal} {J. Phys. A: Math. Theor.}\ }\textbf {\bibinfo {volume} {55}},\ \bibinfo {pages} {225302} (\bibinfo {year} {2022})}\BibitemShut {NoStop}%
\bibitem [{\citenamefont {Colla}\ \emph {et~al.}(2025{\natexlab{a}})\citenamefont {Colla}, \citenamefont {Breuer},\ and\ \citenamefont {Gasbarri}}]{Colla2025}%
  \BibitemOpen
  \bibfield  {author} {\bibinfo {author} {\bibfnamefont {A.}~\bibnamefont {Colla}}, \bibinfo {author} {\bibfnamefont {H.-P.}\ \bibnamefont {Breuer}}, \ and\ \bibinfo {author} {\bibfnamefont {G.}~\bibnamefont {Gasbarri}},\ }\href {https://arxiv.org/abs/2506.04097} {\enquote {\bibinfo {title} {Unveiling coherent dynamics in non-markovian open quantum systems: exact expression and recursive perturbation expansion},}\ } (\bibinfo {year} {2025}{\natexlab{a}}),\ \Eprint {http://arxiv.org/abs/2506.04097} {arXiv:2506.04097 [quant-ph]} \BibitemShut {NoStop}%
\bibitem [{\citenamefont {Alicki}(1979)}]{Alicki1979May}%
  \BibitemOpen
  \bibfield  {author} {\bibinfo {author} {\bibfnamefont {R.}~\bibnamefont {Alicki}},\ }\href {\doibase 10.1088/0305-4470/12/5/007} {\bibfield  {journal} {\bibinfo  {journal} {J. Phys. A: Math. Gen.}\ }\textbf {\bibinfo {volume} {12}},\ \bibinfo {pages} {L103} (\bibinfo {year} {1979})}\BibitemShut {NoStop}%
\bibitem [{\citenamefont {Colla}\ \emph {et~al.}(2025{\natexlab{b}})\citenamefont {Colla}, \citenamefont {Hasse}, \citenamefont {Palani}, \citenamefont {Schaetz}, \citenamefont {Breuer},\ and\ \citenamefont {Warring}}]{Colla2025-ne}%
  \BibitemOpen
  \bibfield  {author} {\bibinfo {author} {\bibfnamefont {A.}~\bibnamefont {Colla}}, \bibinfo {author} {\bibfnamefont {F.}~\bibnamefont {Hasse}}, \bibinfo {author} {\bibfnamefont {D.}~\bibnamefont {Palani}}, \bibinfo {author} {\bibfnamefont {T.}~\bibnamefont {Schaetz}}, \bibinfo {author} {\bibfnamefont {H.-P.}\ \bibnamefont {Breuer}}, \ and\ \bibinfo {author} {\bibfnamefont {U.}~\bibnamefont {Warring}},\ }\href@noop {} {\bibfield  {journal} {\bibinfo  {journal} {Nat. Commun.}\ }\textbf {\bibinfo {volume} {16}},\ \bibinfo {pages} {2502} (\bibinfo {year} {2025}{\natexlab{b}})}\BibitemShut {NoStop}%
\bibitem [{\citenamefont {Davies}(1974)}]{Davies1974}%
  \BibitemOpen
  \bibfield  {author} {\bibinfo {author} {\bibfnamefont {E.}~\bibnamefont {Davies}},\ }\href {\doibase 10.1007/BF01608389} {\bibfield  {journal} {\bibinfo  {journal} {Commun.Marh.Phys.}\ }\textbf {\bibinfo {volume} {39}},\ \bibinfo {pages} {91} (\bibinfo {year} {1974})}\BibitemShut {NoStop}%
\bibitem [{\citenamefont {Colla}\ and\ \citenamefont {Breuer}(2024)}]{Colla2024Dec}%
  \BibitemOpen
  \bibfield  {author} {\bibinfo {author} {\bibfnamefont {A.}~\bibnamefont {Colla}}\ and\ \bibinfo {author} {\bibfnamefont {H.-P.}\ \bibnamefont {Breuer}},\ }\href {\doibase 10.1088/2058-9565/ad98be} {\bibfield  {journal} {\bibinfo  {journal} {Quantum Sci. Technol.}\ }\textbf {\bibinfo {volume} {10}},\ \bibinfo {pages} {015047} (\bibinfo {year} {2024})}\BibitemShut {NoStop}%
\bibitem [{\citenamefont {Chruściński}(2022)}]{Chruscinski2022}%
  \BibitemOpen
  \bibfield  {author} {\bibinfo {author} {\bibfnamefont {D.}~\bibnamefont {Chruściński}},\ }\href {\doibase 10.1016/j.physrep.2022.09.003} {\bibfield  {journal} {\bibinfo  {journal} {Physics Reports}\ }\textbf {\bibinfo {volume} {992}},\ \bibinfo {pages} {1} (\bibinfo {year} {2022})}\BibitemShut {NoStop}%
\bibitem [{\citenamefont {Morozov}\ and\ \citenamefont {Röpke}(2012)}]{Morozov_2012}%
  \BibitemOpen
  \bibfield  {author} {\bibinfo {author} {\bibnamefont {Morozov}}\ and\ \bibinfo {author} {\bibnamefont {Röpke}},\ }\href {\doibase 10.5488/cmp.15.43004} {\bibfield  {journal} {\bibinfo  {journal} {Condensed Matter Physics}\ }\textbf {\bibinfo {volume} {15}},\ \bibinfo {pages} {43004} (\bibinfo {year} {2012})}\BibitemShut {NoStop}%
\bibitem [{\citenamefont {Spohn}\ and\ \citenamefont {Lebowitz}(1978)}]{Lebowitz1978}%
  \BibitemOpen
  \bibfield  {author} {\bibinfo {author} {\bibfnamefont {H.}~\bibnamefont {Spohn}}\ and\ \bibinfo {author} {\bibfnamefont {J.~L.}\ \bibnamefont {Lebowitz}},\ }\enquote {\bibinfo {title} {Irreversible thermodynamics for quantum systems weakly coupled to thermal reservoirs},}\ in\ \href {\doibase https://doi.org/10.1002/9780470142578.ch2} {\emph {\bibinfo {booktitle} {Advances in Chemical Physics}}}\ (\bibinfo  {publisher} {John Wiley \& Sons, Ltd},\ \bibinfo {year} {1978})\ pp.\ \bibinfo {pages} {109--142}\BibitemShut {NoStop}%
\bibitem [{\citenamefont {Serafini}(2017)}]{Serafini2017Jul}%
  \BibitemOpen
  \bibfield  {author} {\bibinfo {author} {\bibfnamefont {A.}~\bibnamefont {Serafini}},\ }\href {\doibase 10.1201/9781315118727} {\emph {\bibinfo {title} {{Quantum Continuous Variables:A Primer of Theoretical Methods}}}}\ (\bibinfo  {publisher} {Taylor {\&} Francis},\ \bibinfo {address} {Andover, England, UK},\ \bibinfo {year} {2017})\BibitemShut {NoStop}%
\bibitem [{\citenamefont {Picatoste}\ \emph {et~al.}(2024)\citenamefont {Picatoste}, \citenamefont {Colla},\ and\ \citenamefont {Breuer}}]{Picatoste2024Mar}%
  \BibitemOpen
  \bibfield  {author} {\bibinfo {author} {\bibfnamefont {I.~A.}\ \bibnamefont {Picatoste}}, \bibinfo {author} {\bibfnamefont {A.}~\bibnamefont {Colla}}, \ and\ \bibinfo {author} {\bibfnamefont {H.-P.}\ \bibnamefont {Breuer}},\ }\href {\doibase 10.1103/PhysRevResearch.6.013258} {\bibfield  {journal} {\bibinfo  {journal} {Phys. Rev. Res.}\ }\textbf {\bibinfo {volume} {6}},\ \bibinfo {pages} {013258} (\bibinfo {year} {2024})}\BibitemShut {NoStop}%
\bibitem [{\citenamefont {Ban}\ \emph {et~al.}(2010)\citenamefont {Ban}, \citenamefont {Kitajima},\ and\ \citenamefont {Shibata}}]{Ban2010}%
  \BibitemOpen
  \bibfield  {author} {\bibinfo {author} {\bibfnamefont {M.}~\bibnamefont {Ban}}, \bibinfo {author} {\bibfnamefont {S.}~\bibnamefont {Kitajima}}, \ and\ \bibinfo {author} {\bibfnamefont {F.}~\bibnamefont {Shibata}},\ }\href@noop {} {\bibfield  {journal} {\bibinfo  {journal} {Phys. Lett. A}\ }\textbf {\bibinfo {volume} {374}},\ \bibinfo {pages} {2324} (\bibinfo {year} {2010})}\BibitemShut {NoStop}%
\bibitem [{\citenamefont {Doll}\ \emph {et~al.}(2008)\citenamefont {Doll}, \citenamefont {Zueco}, \citenamefont {Wubs}, \citenamefont {Kohler},\ and\ \citenamefont {H\"anggi}}]{Doll2008}%
  \BibitemOpen
  \bibfield  {author} {\bibinfo {author} {\bibfnamefont {R.}~\bibnamefont {Doll}}, \bibinfo {author} {\bibfnamefont {D.}~\bibnamefont {Zueco}}, \bibinfo {author} {\bibfnamefont {M.}~\bibnamefont {Wubs}}, \bibinfo {author} {\bibfnamefont {S.}~\bibnamefont {Kohler}}, \ and\ \bibinfo {author} {\bibfnamefont {P.}~\bibnamefont {H\"anggi}},\ }\href {\doibase https://doi.org/10.1016/j.chemphys.2007.09.003} {\bibfield  {journal} {\bibinfo  {journal} {Chemical Physics}\ }\textbf {\bibinfo {volume} {347}},\ \bibinfo {pages} {243} (\bibinfo {year} {2008})}\BibitemShut {NoStop}%
\bibitem [{\citenamefont {Wi{\ss}mann}(2016)}]{Wissmann2016}%
  \BibitemOpen
  \bibfield  {author} {\bibinfo {author} {\bibfnamefont {S.}~\bibnamefont {Wi{\ss}mann}},\ }\href@noop {} {\enquote {\bibinfo {title} {Non-markovian quantum probes for complex systems},}\ }\bibinfo {howpublished} {\url{https://freidok.uni-freiburg.de/data/11335}} (\bibinfo {year} {2016})\BibitemShut {NoStop}%
\end{thebibliography}%

\end{document}